\documentclass[preprint]{JHEP3}

\usepackage{epsfig,multicol,multirow}
\usepackage{amsmath,subfigure,latexsym,amssymb,cite}

\def\lsi{\raise0.3ex\hbox{$<$\kern-0.75em\raise-1.1ex\hbox{$\sim$}}}
\def\gsi{\raise0.3ex\hbox{$>$\kern-0.75em\raise-1.1ex\hbox{$\sim$}}}

\newcommand {\tr}{{\rm tr\,}}

\def\bb#1{\hbox{\mybbs#1}}
\font\mybb=msbm10 at 12pt
\def\bb#1{\hbox{\mybb#1}}

\def\IR{{\bb R}}

\newcommand {\beq}{\begin{equation}}
\newcommand {\eeq}{\end{equation}}
\newcommand {\beqa}{\begin{eqnarray}}
\newcommand {\eeqa}{\end{eqnarray}}

\newcommand {\ee}{\mbox{e}}

\newcommand\fverb{\setbox\pippobox=\hbox\bgroup\verb}
\newcommand\fverbdo{\egroup\medskip\noindent%
                        \fbox{\unhbox\pippobox}\ }
\newcommand\fverbit{\egroup\item[\fbox{\unhbox\pippobox}]}
\newbox\pippobox

\preprint{
HU-EP-06/16\\ 
KEK-TH-1079\\
UTEP-517}

\title{
A non-perturbative study of
4d U(1) non-commutative gauge theory 
--- the fate of one-loop instability
}

\author{
Wolfgang Bietenholz${}^{a}$,
Jun Nishimura${}^{bc}$,
Yoshiaki Susaki${}^{bd}$ and
Jan Volkholz${}^{a}$\\
\llap{$^a$}Institut f\"{u}r Physik, Humboldt Universit\"{a}t zu Berlin \\
        Newtonstr. 15, D-12489 Berlin, Germany \\
\llap{$^b$}High Energy Accelerator Research Organization (KEK), \\
Tsukuba, Ibaraki, 305-0801, Japan \\
\llap{$^c$}Department of Particle and Nuclear Physics,\\
Graduate University for Advanced Studies (SOKENDAI),\\
Tsukuba, Ibaraki 305-0801, Japan \\
\llap{$^d$}Graduate School of Pure and Applied Science,
University of Tsukuba,\\
Tsukuba, Ibaraki 305-8571, Japan\\
\email{bietenho@physik.hu-berlin.de,
jnishi@post.kek.jp,
susaki@post.kek.jp,
volkholz@physik.hu-berlin.de}} 

\abstract{
Recent perturbative studies show that
in 4d non-commutative spaces,
the trivial (classically stable) 
vacuum of gauge theories becomes unstable at the quantum level,
unless one introduces sufficiently many fermionic degrees of freedom.
This is due to a negative IR-singular term in the one-loop effective
potential, which appears as a result of the UV/IR mixing.
We study such a system non-perturbatively
in the case of pure U(1) gauge theory in four dimensions,
where two directions are non-commutative.
Monte Carlo simulations are performed
after mapping the regularized theory 
onto a U($N$) lattice gauge theory in $d=2$.
At intermediate coupling strength,
we find a phase in which open Wilson lines acquire non-zero 
vacuum expectation values,
which implies the spontaneous breakdown of translational invariance.
In this phase, various physical quantities
obey clear scaling behaviors
in the continuum limit with a fixed non-commutativity parameter $\theta$,
which provides evidence for a possible continuum theory.
The extent of the dynamically generated space 
in the non-commutative directions
becomes finite 
in the above limit, and its dependence on $\theta$
is evaluated explicitly.
We also study the dispersion relation.
In the weak coupling symmetric phase, 
it involves a negative IR-singular term, 
which is responsible for the observed
phase transition. In the broken phase, it reveals
the existence of the Nambu-Goldstone mode associated with
the spontaneous symmetry breaking.
}


\begin{document}

\section{Introduction}

Non-commutative (NC) geometry \cite{Sny,Connes}
has been studied extensively
as a modification of our notion of space-time
at short distances, possibly due to effects of 
quantum gravity \cite{gravity}.
It has recently attracted much attention since gauge theories
on a NC geometry have been shown to appear as
a low energy limit of string theories with a background 
tensor field \cite{String}.
At the classical level, introducing non-commutativity to
the space-time coordinates modifies the ultraviolet dynamics
of field theories (the interaction becomes non-local at the
scale of non-commutativity),
but not the infrared properties.
The latter changes at the quantum level, however, 
due to the so-called UV/IR mixing effect \cite{rf:MRS}.
As a consequence,
one cannot retrieve the
corresponding commutative theory by sending the non-commutativity
parameter to zero in general.

The UV/IR mixing poses a severe problem 
in the renormalization procedure
within perturbation theory
since a new type of IR divergences, 
in addition to the usual UV divergences \cite{UV},
appears in non-planar diagrams \cite{rf:MRS}.
%
The finite lattice formulation \cite{AMNS} based on twisted reduced models
\cite{GAO} and their new interpretation \cite{AIIKKT} in the context
of NC geometry, regularizes such divergences
--- not only the ordinary UV divergences but also the novel
IR divergences.
It therefore provides a non-perturbative framework to establish
the existence of a consistent field theory on a NC geometry.
In ref.\ \cite{2dU1} a simple field theory,
2d U(1) gauge theory on a NC 
plane\footnote{See refs.\ \cite{Mafia,Griguolo:2003kq} 
for related analytic works,
including several attempts to actually solve this model.},
has been studied by Monte Carlo simulation, and the existence of a
finite continuum limit has been confirmed.
The ultraviolet dynamics is described by the commutative
2d U($\infty$) gauge theory.
On the other hand, the infrared dynamics would be described by
the commutative 2d U(1) theory, were it not for the UV/IR mixing effects.
The ``dynamical Aharonov-Bohm effect'' observed in this regime
clearly demonstrates the non-equivalence to the commutative 2d U(1) theory.

As an interesting physical consequence of the UV/IR mixing
in the case of scalar field theory,
ref.\ \cite{GuSo} predicted the existence of a ``striped phase'',
in which non-zero Fourier modes of the scalar field
acquire a vacuum expectation value
and break the translational invariance spontaneously.
This prediction, which was based on a self-consistent 
Hartree-Fock approximation, was supported later by 
another study using an effective action \cite{CZ},
and discussed also in the framework of a renormalization group 
analysis in $4-\varepsilon$ dimensions \cite{ChenWu}.
Finally, 
the existence of this new phase was fully established
by Monte Carlo simulations
\cite{Procs,AC,Martin:2004un,Bietenholz:2004xs,Julieta}.
In ref.\ \cite{Bietenholz:2004xs},
in particular, the explicit results for the phase diagram 
and the dispersion relation were presented in $d=3$.
The dispersion relation in the disordered phase shows 
a positive IR singularity due to the UV/IR mixing, 
which shifts the energy minimum to a non-vanishing momentum.
As one changes the mass parameter in the action,
the mode at the energy minimum condenses, which 
yields
the corresponding stripe pattern. 
The existence of a sensible continuum limit is suggested 
by the scaling of various physical quantities.
In particular the average width of the stripes
stabilizes at a finite value
in the continuum limit.
In even dimensions with the non-commutativity tensor of maximal rank,
the emergence of the striped phase can also be conjectured
from the eigenvalue distribution of the matrix, which
represents the scalar field in NC geometry
\cite{Martin:2004un,Steinacker:2005wj}.

In the case of 4d U(1) NC gauge theories,
the IR singularity occurs
in the one-loop calculation of 
the vacuum polarization tensor \cite{hayakawa}.
This changes the dispersion relation of photons 
at low momenta \cite{mst}.
In fact the IR singularity in the dispersion relation
occurs with an overall {\em negative} 
sign \cite{LLT,ruiz,Bassetto:2001vf}
unless one introduces sufficiently many fermionic degrees of freedom.
%
It turns out that
the manifestly gauge invariant effective action, which is derived
from field theoretical calculations
as well as from string theoretical calculations \cite{rf:MVR,rf:AL},
involves the open Wilson line operators \cite{IIKK},
which implies that the IR singularity is associated 
with their condensation.
%
This causes the spontaneous breakdown of translational invariance
since the open Wilson lines carry specific non-zero momenta
dictated by the star-gauge invariance \cite{IIKK}.
In the case of a finite NC torus,
the same phenomenon has been understood \cite{Guralnik:2002ru}
from its Morita dual {\em commutative} SU($N$) gauge theory 
with twisted boundary conditions\footnote{At high tempertature 
(on the scale of the non-commutativity)
the ring diagrams dominate. Their evaluation to
all orders implies a critical temperature, above which the
magnetic photon mass squared turns negative in the NC
planes \cite{BFM}.}.

We should note, however, that these perturbative analyses merely imply
that the {\em perturbative vacuum} of 
4d NC gauge theory has tachyonic instability,
but they do not exclude the possibility
that the system eventually finds
a stable vacuum. (An analogous scenario known as
the ``tachyon condensation'' has been proposed in the
open string field theory \cite{tachyon-condensation}\footnote{This 
might be more than a plain analogy since the tachyonic instability
of NC gauge theory has partial contribution
from the closed string tachyon \cite{rf:AL,Seiberg:2000zk}.}.)
In order to investigate this issue, we definitely need to employ
a non-perturbative framework.

In the current work, 
the lattice formulation \cite{AMNS} is particularly
important since it allows us to study the model from first principles
by Monte Carlo simulation.
We study pure U(1) gauge theory in four dimensions,
where two directions are non-commutative,
and we find a phase in which open Wilson lines acquire non-zero 
vacuum expectation values.
In fact this phase extends towards weak coupling as the system
size is increased, and it turns out that we always
end up in this phase in the continuum and infinite-volume limits
with a fixed non-commutativity parameter.
This is consistent with the prediction from the one-loop calculations.
We observe,
however, that various physical quantities
obey clear scaling behaviors in the above limit,
which provides evidence for a possible continuum theory.
We also study the dispersion relation.
In the weak coupling symmetric phase, it involves a negative IR-singular term,
which makes the energy
vanish at a non-zero momentum as one approaches the critical point.
In the broken phase, although the momentum components
in the NC directions are no longer conserved, 
we can still study the
relation between the energy and the momentum in the {\em commutative}
direction. This reveals
the existence of the Nambu-Goldstone mode associated with
the spontaneous symmetry breakdown.

This paper is organized as follows.
In section \ref{sec:review} we define
a gauge theory in 4d NC space,
and we briefly review the results obtained by perturbative calculations.
In section \ref{sec:lattice} we introduce the lattice formulation
and discuss how to take the continuum limit.
In section \ref{sec:phase} we present the phase diagram
of the lattice model.
In section \ref{sec:double-scaling} we investigate
the existence
of a sensible continuum limit.
In section \ref{sec:phys-extent-NC} we study how the
space in the NC directions 
looks like in the broken phase.
In section \ref{sec:dispersion-rel} we study the dispersion
relation in each phase to gain deeper understanding of the
observed phase transition and the continuum limit.
Section \ref{sec:summary} is devoted to a summary and discussions.
In the appendix we explain the algorithm used for our Monte Carlo
simulations.
Some preliminary results of this work have been presented
in proceeding contributions \cite{proc}.

\section{Tachyonic instability in perturbation theory} 
\label{sec:review}

NC geometry is characterized by the following commutation
relation among the space-time coordinates
\beq
[ \hat{x}_\mu , \hat{x}_\nu] = i \, \Theta_{\mu \nu} \ ,
\eeq
where $\Theta_{\mu \nu}$ is the non-commutativity tensor.
Here we consider the 4d Euclidean
space-time $\hat{x}_{\mu}$ ($\mu=1,\cdots , 4$)
with the non-commutativity introduced only in the $\mu=1,2$ directions,
i.e.,
\beqa
\Theta_{12} &=& - \Theta_{21} = \theta \ , \nonumber \\
\Theta_{\mu\nu} &=& 0 \quad \quad \mbox{otherwise} \ .
\label{def-Theta}
\eeqa
Since we have two commutative directions, we may regard one of them
as the Euclidean time.
This allows us to alleviate the well-known problems 
concerning causality and unitarity \cite{causal,unitar}.
Moreover, it also 
enables us to define the dispersion relation as a useful probe
to study the system \cite{Bietenholz:2004xs}.

Pure U(1) gauge theory in this NC space
can be formulated in the path integral formalism by the gauge action
\begin{eqnarray}
S &=& \frac{1}{4} \int d^{4}x  \,  F_{\mu \nu}(x)
\star F_{\mu \nu}(x) \ ,
\nonumber \\
F_{\mu \nu} &=& \partial_{\mu} A_{\nu} - \partial_{\nu}A_{\mu}
+ i \, g \, (A_{\mu} \star A_{\nu} - A_{\nu} \star A_{\mu} ) \ , 
\label{actNC}
\end{eqnarray}
where the space-time coordinates $x_\mu$ are c-numbers as
in ordinary field theories, but
the non-commutativity is now encoded in the star product
\beq
f(x) \star g(x) = \left. \exp \left( \frac{i}{2} \Theta_{\mu \nu}
\frac{\partial}{\partial x_{\mu}} \frac{\partial}{\partial y_{\nu}} \right)
f(x) g(y) \right|_{x=y}  \ .
\label{def-starprod}
\eeq
The action in eq.~(\ref{actNC}) is invariant
under a star-gauge transformation
\beq
\label{gauge}
A_\mu \mapsto A_\mu + 
\partial_\mu \Lambda + ig[A_\mu, \Lambda]_\star \ , 
\eeq
where the star-commutator is defined by
$[f,g]_\star \equiv f\star g - g\star f$.
Note that non-linear terms appear 
in the field strength tensor in eq.\ (\ref{actNC})
due to the NC geometry, although we are dealing with a gauge group
of rank 1. As a result, the theory shares such 
properties\footnote{It is also conjectured that NC photons
form bound states analogous to glueballs in QCD \cite{Fatollahi:2005ri}.}
as a negative beta function \cite{hayakawa,ruiz}
with non-Abelian (rather than Abelian) 
gauge theories in the commutative space.

This model has been studied extensively in perturbation theory.
In particular the result that is most relevant for us
is the one-loop effective action for the gauge field,
which involves the quadratic term
\cite{hayakawa,mst,LLT,ruiz,Bassetto:2001vf}
\beqa
\label{ea}
\Gamma_{\rm 1-loop} &=& 
  {g^2 \over \pi^{2} } 
\int {d^4 p \over (2 \pi)^4} 
A_\mu(p) A_\nu(-p) 
{\tilde{p}_\mu \tilde{p}_\nu \over |\tilde{p}|^4} + \cdots \ , \\
 \mbox{where} \quad \tilde{p}_\mu &=& \Theta_{\mu \nu} p_\nu \ .
\nonumber 
\eeqa
This term emerges from the non-planar diagrams and represents
an effect of UV/IR mixing.
Since the effective potential, which is 
{\em minus} the effective action according to the adopted convention,
involves a {\em negative} quadratic term,
we find that the low momentum modes of the gauge field cause
a tachyonic instability.

In fact the quadratic term in (\ref{ea}) is invariant
under the star-gauge transformation (\ref{gauge})
only at the leading order in $A_\mu$.
In order to obtain a fully gauge invariant effective action,
one needs to include terms
at higher orders in $A_\mu$ as in the supersymmetric case \cite{lm}.
The term in the gauge-invariant effective action 
which is most singular at small $p$ is given by 
\cite{rf:MVR,rf:AL}\footnote{
Refs.\ \cite{SJR} pointed out that
the appearance of open Wilson lines in the effective action
is a generic feature of field theories on NC geometry.
}
\beq
\label{density}
\Gamma_{\rm 1-loop} = 
{ 1 \over \pi^2 }
\int {d^4 p \over (2 \pi)^4} \, 
W(p) W(-p) {1 \over |\tilde{p}|^4}  \ .
\eeq
The open Wilson line $W(p)$, which appears here, 
is a manifestly
star-gauge invariant operator defined as \cite{IIKK}
\beq
\label{wilson}
W(p) = \int d^d x \, \ee^{i p \cdot x} \, 
{\cal P} \exp _\star
\left( i \, 
g \int_x^{x+\tilde{p}} A_\mu(\xi ) \, d \xi_\mu \right)  \ ,
\eeq
where ${\cal P} \exp$ represents the path-ordered exponential,
and the path for the line integral over $\xi$ is taken to 
be a straight line connecting $x$ and $x+\tilde{p}$.
Expanding (\ref{density}) in terms of $A_\mu$, one obtains
(\ref{ea}) at the leading order, up to an irrelevant constant term.
The result (\ref{density}) implies that the instability
is associated with 
the condensation of open Wilson lines $W(p)$ with small $p$.
Since the open Wilson line $W(p)$ carries non-zero momentum,
its condensation causes the spontaneous breakdown of
translational symmetry.

{}From the perturbative analysis alone, one cannot tell whether
the theory possesses a stable non-perturbative vacuum after
the condensation of open Wilson lines.
To answer this question and to study the nature of the theory
at the stable vacuum, if it exists,
we definitely need a fully non-perturbative approach.

\section{The lattice model and its continuum limit}
\label{sec:lattice}

\subsection{Lattice regularization of NC gauge theory}

The lattice regularized version of 
the theory (\ref{actNC}) can be defined by an
analog of Wilson's plaquette action \cite{AMNS}
\beq
S= - \beta \sum_{x} \sum_{\mu < \nu}
U_\mu (x) \star U_\nu (x + a \hat{\mu}) \star
U_\mu (x + a \hat{\nu})^\ast \star U_\nu (x)^\ast
 + \mbox{c.c.} \ ,
\label{lat-action}
\eeq
where 
the symbol $\hat{\mu}$ represents 
a unit vector in the $\mu$-direction
and we have introduced the lattice spacing $a$.
The link variables $U_\mu(x)$ ($\mu = 1, \cdots , 4$)
are complex fields on the lattice
satisfying the star-unitarity condition
\beq
U_\mu(x) \star U_\mu(x)^{\ast} = 
 U_\mu(x)^{\ast} \star U_\mu(x) = 1  \ .
\label{star-unitary}
\eeq
The star product on the lattice can be obtained by
rewriting (\ref{def-starprod}) in terms of Fourier modes
and restricting the momenta to the Brillouin zone.
The action (\ref{lat-action})
is invariant under a star-gauge transformation
\beq
U_\mu(x) \mapsto 
g(x)\star U_\mu(x) \star g (x+ a \hat{\mu}) ^\ast  \ ,
\label{star-gauge-lat}
\eeq 
where also $g(x)$ obeys
the star-unitarity condition
\beq
g(x) \star g(x)^{*} = 
 g(x)^{*} \star g(x) = 1 \ .
\label{star-unitaryg}
\eeq
As in the commutative space,
one obtains the continuum action (\ref{actNC})
from (\ref{lat-action}) in the $a \rightarrow 0$ limit
with the identification $\beta = \frac{1}{2 g^2}$ and
\beq
U_\mu(x) = 
{\cal P} \exp _\star
\left( i\, g \int_x^{x+a \hat{\mu}} 
A_\mu(\xi ) \, d \xi_\mu \right)  \ .
%
\eeq

In order to study the lattice NC theory (\ref{lat-action})
by Monte Carlo simulations,
it is crucial to reformulate
it in terms of matrices \cite{AMNS}.
In the present setup (\ref{def-Theta}), with two NC directions
and two commutative ones,
the transcription applies only to
the NC directions, and the commutative directions remain untouched.
Let us decompose the four-dimensional coordinate as 
$x\equiv (y,z)$, where 
\beq
y\equiv (x_1 , x_2) \, \quad \mbox{and} \quad z\equiv (x_3 , x_4)
\eeq
represent two-dimensional coordinates 
in the NC and in the commutative plane, respectively.
We use a one-to-one map between a field $\varphi(x)$
on the four-dimensional $N \times N \times L \times L$ lattice 
and a $N \times N$ matrix field $\hat{\varphi}(z)$ on a two-dimensional 
$L \times L$ lattice.
This map yields the following correspondence
\beqa
\varphi_1 (y,z) \star \varphi_2 (y,z) \quad &\Longleftrightarrow& \quad
\hat{\varphi}_1(z) \, \hat{\varphi}_2(z)  \ , \\
\varphi (y+a\hat{\mu},z) \quad &\Longleftrightarrow& \quad
\Gamma_\mu \, \hat{\varphi}(z) \, \Gamma_\mu^\dag \ , \\
\frac{1}{N^2} \sum_{y} \varphi (y,z) \quad &\Longleftrightarrow& \quad
\frac{1}{N}\, \tr \hat{\varphi}(z) \ .
\eeqa
The SU($N$) matrices $\Gamma_\mu$ ($\mu = 1, 2$),
which represent a shift in a NC direction,
satisfy the 't Hooft-Weyl algebra
\beqa
\label{tH-W-alg}
\Gamma_1 \Gamma_2 &=& {\cal  Z}_{12} \, 
\Gamma_2 \Gamma_1   \ , \\
{\cal  Z}_{12}&=&{\cal  Z}_{21}^* = 
\exp\left( \pi i \, \frac{N+1}{N} \right)  \ ,
\label{def-twist}
\eeqa
with the matrix size $N$ being an odd integer.
For this particular construction 
\cite{Nishimura:2001dq,2dU1,Bietenholz:2004xs},
which we are going to use throughout this paper, it turns out that
the non-commutativity parameter $\theta$
in (\ref{def-Theta}) is given by
\beq
\theta = \frac{1}{\pi} N a^2 \ .
\label{theta-def}
\eeq
Note that the extent in the NC directions
$N a$ goes to infinity
in the continuum limit $a \rightarrow 0$ at fixed $\theta$.

Using this map, the link variables $U_\mu(x)$ are mapped
to a $N\times N$ matrix field $\hat{U}_\mu(z)$ on the two-dimensional
$L\times L$ lattice, and the star-unitarity condition (\ref{star-unitary})
simply requires $\hat{U}_\mu(z)$ to be unitary.
The action (\ref{lat-action}) can be rewritten in terms
of the unitary matrix field $\hat{U}_\mu(z)$ in a straightforward manner.
By performing a field redefinition 
\beq
V_\mu(z) \equiv \left\{ 
\begin{array}{ll}
  \hat{U}_\mu(z) \Gamma_\mu
 \quad \quad
&\mbox{for~}\mu=1,2 \ , \\
  \hat{U}_\mu(z)
 \quad \quad
&\mbox{for~}\mu=3,4 \ ,
\end{array}
\right.
\label{Vfield-redef}
\eeq
%
where $V_\mu(z)\in {\rm U}(N)$, we arrive at
\beqa
S &=& S_{\rm NC} + S_{\rm com} + S_{\rm mixed} \ , \nonumber \\
S_{\rm NC} &=& 
- N \beta {\cal Z}_{12} \sum_{z} \, \tr \, 
\Bigr( \,  V_1 (z) \, V_2 (z) \, 
V_1 (z)^\dag \, V_2 (z)^\dag \Bigl)
+ \ \mbox{c.c.}  \ ,  \nonumber \\
S_{\rm com} &=& 
- N \beta  \sum_{z} \, \tr  \, 
\Bigr( \, V_3 (z) \, V_4 (z + a \hat{3}) \, 
V_3 (z + a \hat{4})^\dag \, V_4 (z)^\dag  \Bigl)
 \ + \ \mbox{c.c.}  \ , \nonumber \\
S_{\rm mixed} &=& 
- N \beta \sum_{z} \sum_{\mu=1}^2 \sum_{\nu=3}^4  \, \tr  
\, \Bigr( \,
V_\mu (z) \, V_\nu (z)\, 
V_\mu (z + a \hat{\nu})^\dag \, V_\nu (z)^\dag \Bigl) \ + \ \mbox{c.c.}  \ .
\label{action}
\eeqa
These terms
represent a plaquette in the NC plane, in the commutative
plane and a sum over the plaquettes in the four mixed planes, 
respectively.
Since the action (\ref{action})
is invariant under the gauge transformation
\beq
V_\mu(z) \mapsto \left\{ 
\begin{array}{ll}
  \hat{g}(z) \, V_\mu(z) \,  \hat{g}(z)^\dag \quad \quad
&\mbox{for~}\mu=1,2 \ ,  \\
 \hat{g}(z) \, V_\mu(z)  \, \hat{g}(z+ a \hat{\mu})^\dag 
\quad \quad &\mbox{for~}\mu=3,4 \ ,
\end{array}
\right.
\label{gauge-trf}
\eeq
where $\hat{g}(z)\in {\rm U}(N)$,
it defines a 2d lattice gauge theory,
in which $V_\mu(z)$ ($\mu=3,4$) correspond to link variables,
and $V_\mu(z)$ ($\mu=1,2$) correspond
to scalar fields in the adjoint representation
with the specific self-coupling given by $S_{\rm NC}$.
Taking into account the field redefinition (\ref{Vfield-redef}),
one easily finds that
the U($N$) gauge symmetry (\ref{gauge-trf}) of the 2d theory
corresponds precisely to the star-gauge invariance 
(\ref{star-gauge-lat}) of the 4d NC theory that we started with.

\subsection{Eguchi-Kawai equivalence in the planar limit}

In fact the lattice model (\ref{action}) 
can be obtained by the twisted dimensional reduction \cite{GAO}
from 4d pure U($N$) lattice gauge theory\footnote{In the 
present case, this amounts to considering
pure U($N$) gauge theory on a $1 \times 1 \times L \times L$ lattice
with the twisted boundary condition \cite{'tHooft:1979uj}
for the 1,2 directions,
and the periodic boundary condition for the 3,4 directions.
The phase factor ${\cal Z}_{12}$ in the first term of (\ref{action})
represents the twist.}.
Historically such a model appeared in the context of 
the Eguchi-Kawai reduction \cite{EK}, which in the present case
provides an explicit relation between
the 2d U($N$) theory (\ref{action}) and the 4d U($N$) theory
(both in the commutative space)
in the large-$N$ limit at fixed $\beta$.
(This limit is usually referred to
as the planar limit since only planar diagrams
survive \cite{'tHooft:1973jz}.)
More specifically, the vacuum expectation value of a Wilson loop
defined in each theory coincides in the above limit,
if
the U(1)$^2$ symmetry 
\beq
V_\mu(z) \mapsto \ee^{i \alpha_\mu} V_\mu(z) 
\quad\quad \mbox{for~}\mu=1,2 
\label{U1-2sym}
\eeq
of the 2d theory (\ref{action}) is not spontaneously broken.
We will see that this condition holds in the weak coupling and the 
strong coupling regions, but it is violated in the intermediate
region. In fact this symmetry (\ref{U1-2sym}) includes
the (discrete) translational symmetry of the 4d NC lattice gauge theory in 
the NC directions as a subgroup 
up to a star-gauge transformation \cite{Bietenholz:2004as}. 
Therefore, the spontaneous breaking of the U(1)$^2$ symmetry
corresponds precisely to the IR instability of the perturbative vacuum
discussed in section \ref{sec:review}.

Let us comment on some known results for
a totally reduced (one-site) model.
The condition for the Eguchi-Kawai equivalence in this case
is that the U(1)$^4$ symmetry is not spontaneously broken.
The twist was introduced in ref.\ \cite{GAO} into the original 
(untwisted) Eguchi-Kawai model \cite{EK}
in order to avoid the spontaneous U(1)$^4$ 
symmetry breaking in the weak coupling 
region\footnote{It has been shown recently 
\cite{Hanada:2006vm}
that in the original (untwisted) Eguchi-Kawai model and
in a more general model,
the U(1)$^4$ symmetry breaking does not occur at once, 
but rather in a sequence
U(1)$^4$$\rightarrow$U(1)$^3$$\rightarrow$U(1)$^2$$\rightarrow$U(1)
$\rightarrow$none. Such a partial breaking of U(1)$^4$ symmetry
was observed earlier in large $N$ gauge theories in a finite box 
\cite{Narayanan:2003fc}.}.
Indeed the condition is satisfied in the strong coupling and 
weak coupling expansions, and early Monte Carlo studies
suggested that it holds, too, at intermediate couplings \cite{GAO}.
However, Ishikawa and Okawa \cite{ishikawa-okawa}
have recently performed Monte Carlo simulations with
much larger $N$, and observed a signal of spontaneous symmetry breaking
at intermediate couplings. This behavior persisted also for choices
of the twists other than the minimal one 
adopted in ref.\ \cite{GAO}.

The Eguchi-Kawai equivalence, if the condition is met,
holds in the large-$N$ limit with a finite lattice spacing $a$.
The continuum limit  $a \to 0$ as a commutative 4d U($\infty$) gauge 
theory can be taken in the next step by sending $\beta \to \infty$.
In our model, however,
the critical $\beta$, below which the symmetry breaking occurs,
is observed to increase as $N^2$, which implies that
the Eguchi-Kawai equivalence does not persist in the continuum limit.
It remains to be seen whether this is the case
even if one takes other options in defining
twisted reduced models (the choice of the twist, partial or total 
reduction, etc.).
An alternative to ensure the validity
of the Eguchi-Kawai equivalence
in the continuum limit is the quenched reduced model \cite{BHN}.
Its continuum version \cite{GK} has been applied recently
\cite{Kawahara:2005an} to a non-lattice regularization of 
Matrix Theory \cite{Banks:1996vh}.
See also refs.\ \cite{Narayanan:2003fc,cont-reduction} 
for recent developments in the large-$N$ reduction in the continuum.

\subsection{Double scaling limit}


As one can see from (\ref{theta-def}),
the planar limit corresponds to $\theta = \infty$
in the context of NC field theory.
In order to take the continuum limit 
as a NC gauge theory with finite $\theta$, 
one has to take the $N \to \infty$ and $\beta \to \infty$ limits
simultaneously, while satisfying (\ref{theta-def}), 
which therefore implies $a \to 0$.
Following the usual terminology in matrix models,
we call it the double scaling limit.
Unlike in the planar limit, non-planar diagrams survive and they
cause the intriguing UV/IR mixing effects.
The inverse coupling constant $\beta$ has to be tuned in such a way that
physical quantities converge.
Whether this is possible or not is precisely the issue
of (non-perturbative) renormalizability,
which we address in the following.

In the 2d NC gauge theory \cite{2dU1}, 
since the U(1)$^2$ symmetry is not spontaneously broken 
throughout the whole coupling region,
the Eguchi-Kawai equivalence holds in the planar limit for all $\beta$.
The tuning of $\beta$ after taking the planar limit 
can therefore be deduced from the exact solution \cite{GW} 
of the 2d U($\infty$) lattice gauge theory.
This tuning of $\beta$ with respect to the lattice spacing $a$
should be used also in the double scaling limit
in order to make the correlation functions 
scale in the UV regime \cite{Konagaya:2005nb}.
Whether the scaling extends to the IR regime or not
is a non-trivial issue, which was answered affirmatively
in ref.\ \cite{2dU1}.
In the present 4d case, since the U(1)$^2$ symmetry 
is spontaneously broken in the coupling region
relevant to the continuum limit,
the tuning of $\beta$ cannot be deduced from the 
results in 4d U($\infty$) lattice gauge theory.
We therefore have to fine-tune $\beta$ in such a way
that the double scaling is optimized.


\section{Phase structure}
\label{sec:phase}

To begin with, 
we investigate the phase structure of the lattice model (\ref{action}).
Throughout this paper, 
we take $L$,
the number of sites in the commutative directions,
to be $L=N\pm 1$ so that it becomes a multiple of four
in order to use four processors on a parallel machine
efficiently.
The small anisotropy of the lattice should be negligible at large $N$.
%

As a standard quantity
we plot the normalized action
(or the average plaquette)
\beq
{\cal E} = - \frac{1}{12 N^2 \beta}{\langle S\rangle}
\label{def-normalized}
\eeq
against $\beta$ for $N=25$ in figure \ref{fig:action}.
We have performed a ``thermal cycle'' by
carrying out simulations at varying $\beta$
with the initial configuration
taken from a configuration thermalized with a slightly larger/smaller 
$\beta$.
There is a gap at $\beta \sim 0.35$ (left),
and
we observe a slight tendency of a possible hysteresis behavior 
at $1.1 \lesssim \beta \lesssim 1.6$ (right)
although the difference between the two branches, even it exists,
is too tiny to be confirmed definitely.

  \FIGURE{
\epsfig{file=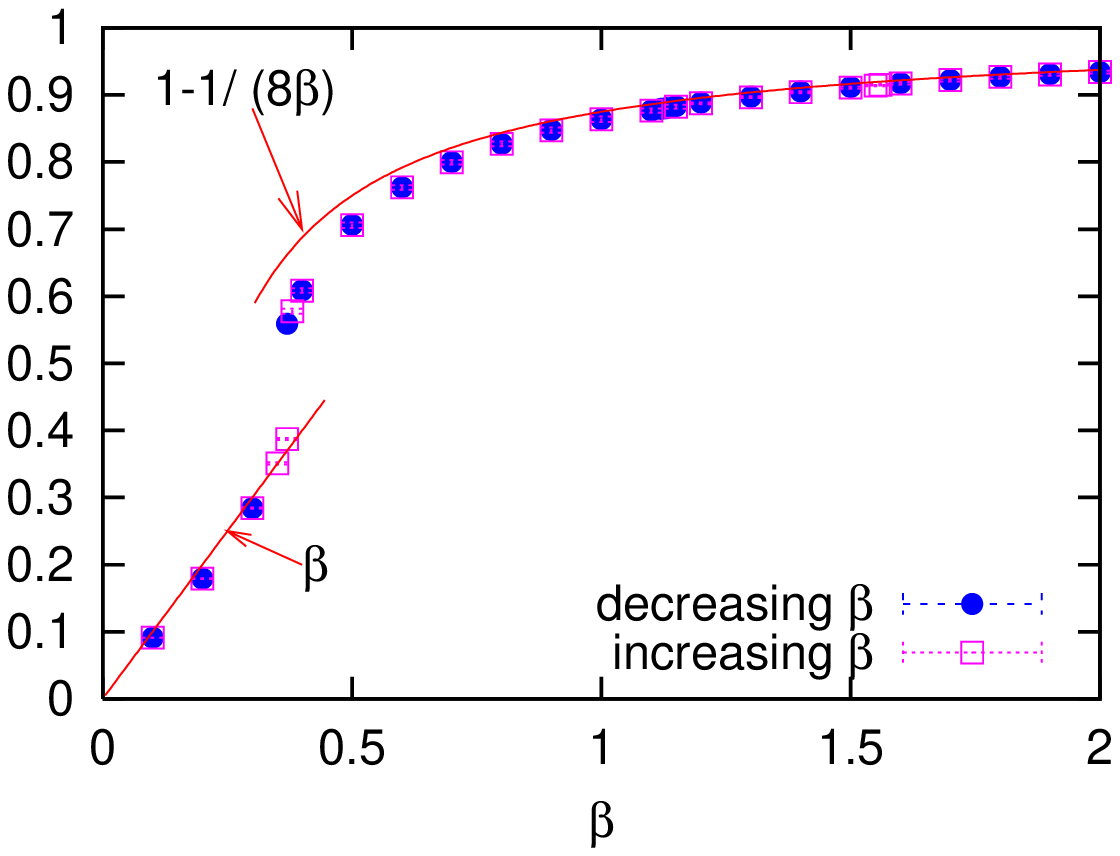,angle=270,width=7.4cm}
\epsfig{file=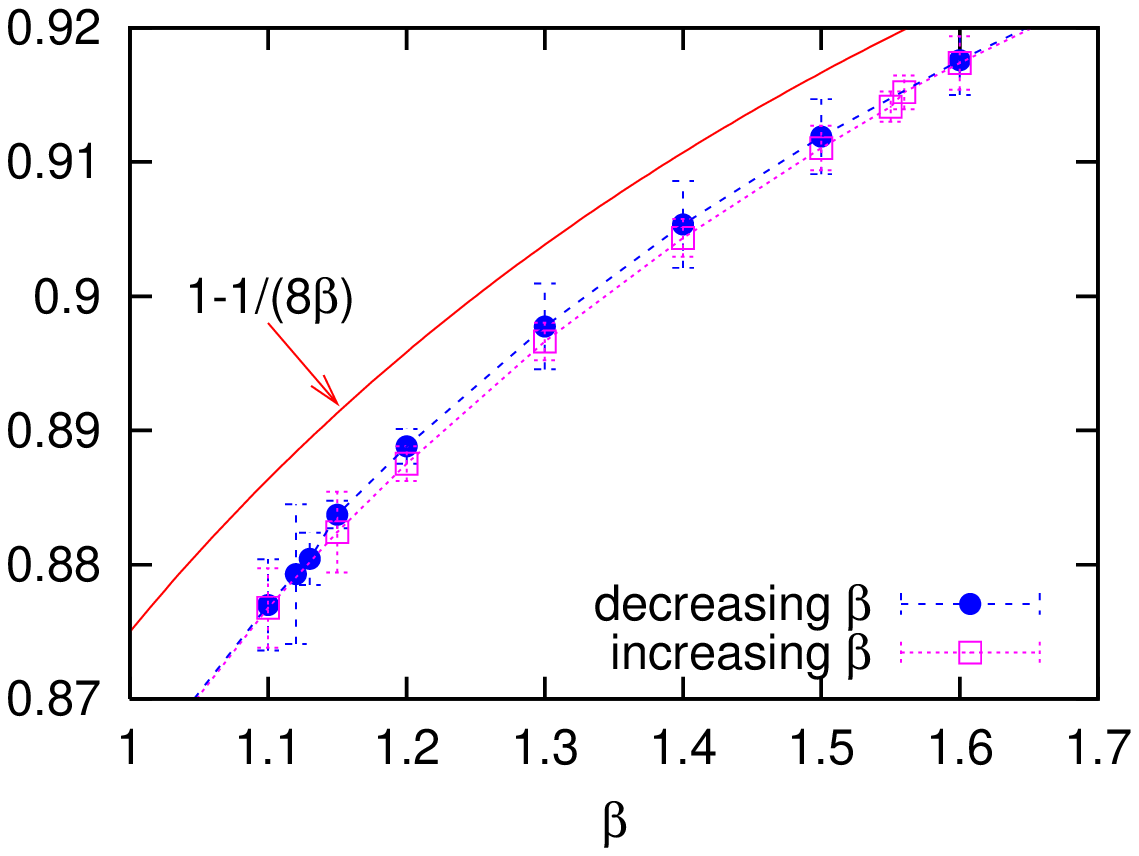,angle=270,width=7.4cm}
   \caption{(Left) The expectation value of the normalized action
(\ref{def-normalized}) is plotted against $\beta$ for $N=25$.
The two kinds of symbols are used to distinguish the results
obtained for decreasing $\beta$ and increasing $\beta$
in the ``thermal cycle''.
The solid lines represent the results of the strong and the
weak coupling expansions.
The discontinuity is observed at $\beta \sim 0.35$.
(Right) The zoom-up of the same plot
around $1.1 \lesssim \beta \lesssim 1.6$.
}
  \label{fig:action}
}

As an order parameter for the spontaneous breaking
of the U(1)$^2$ symmetry (\ref{U1-2sym}), we define 
a gauge invariant operator
\begin{equation}
P_\mu(n) = 
\frac{1}{N L^2} \sum_z
\tr 
\Bigl( V_\mu(z)^n \Bigr) 
~~~\mbox{for}~ \mu=1,2 \ ,\\ 
\label{orderparameter}
\end{equation}
which transforms non-trivially under (\ref{U1-2sym}).
This operator corresponds to the open Wilson line (\ref{wilson})
carrying a momentum with the absolute value \cite{AMNS,2dU1}
\beq
p=\frac{2\pi k}{Na} \ , \quad \quad \quad
k=\left\{\begin{array}{cc}
\frac{n}{2}~~~\mbox{for even}~ n \ , \\ 
\frac{n+N}{2}~~\mbox{for odd}~ n  \ .
\end{array}\right.
\label{momentum-polyakov}
\eeq
Since the operator $P_\mu(n)$ with odd $n$ carries a momentum
of the cutoff order, it does not couple to excitations
that survive in the continuum limit\footnote{This Z$_2$ grading 
of the open Wilson line operators 
should be regarded as an artifact of the 
lattice regularization \cite{AMNS}.}.
Therefore we will focus mainly on the even $n$ case in what follows.

  \FIGURE{
\epsfig{file=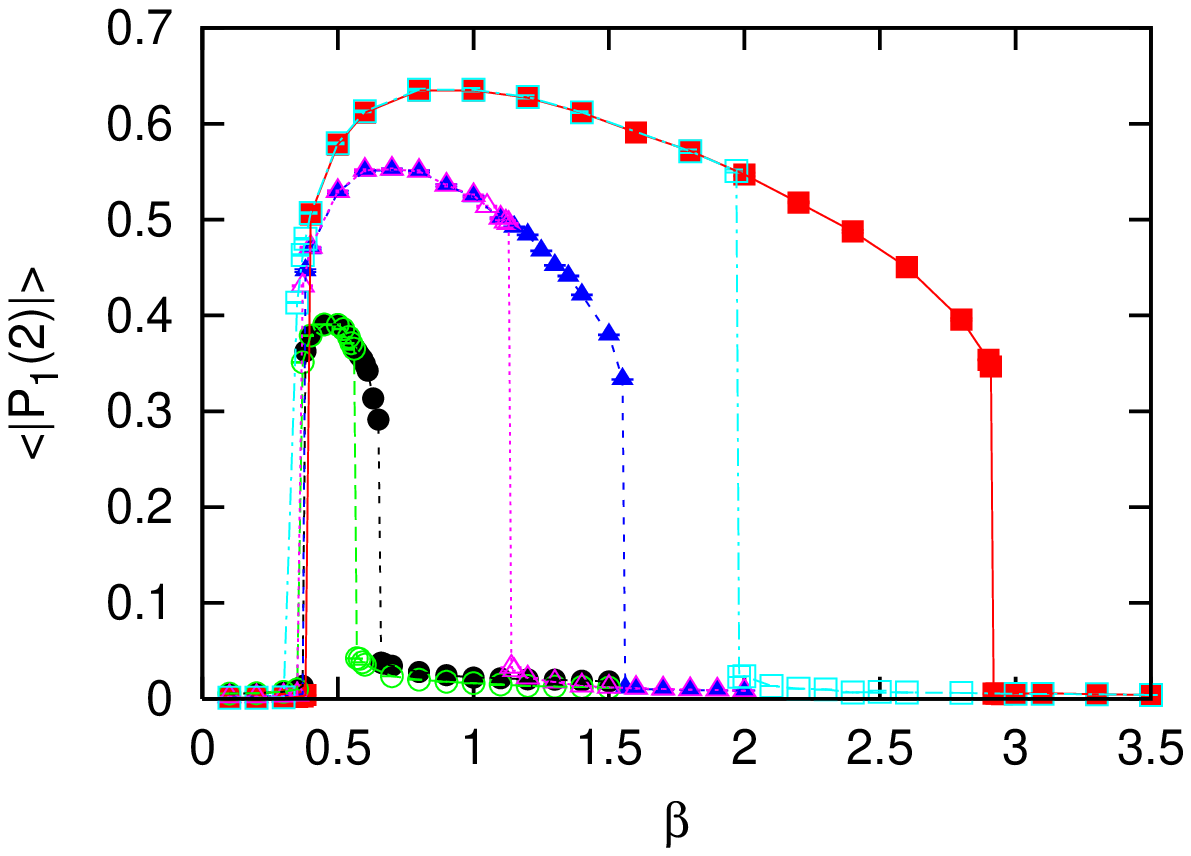,angle=270,width=7.4cm}
\epsfig{file=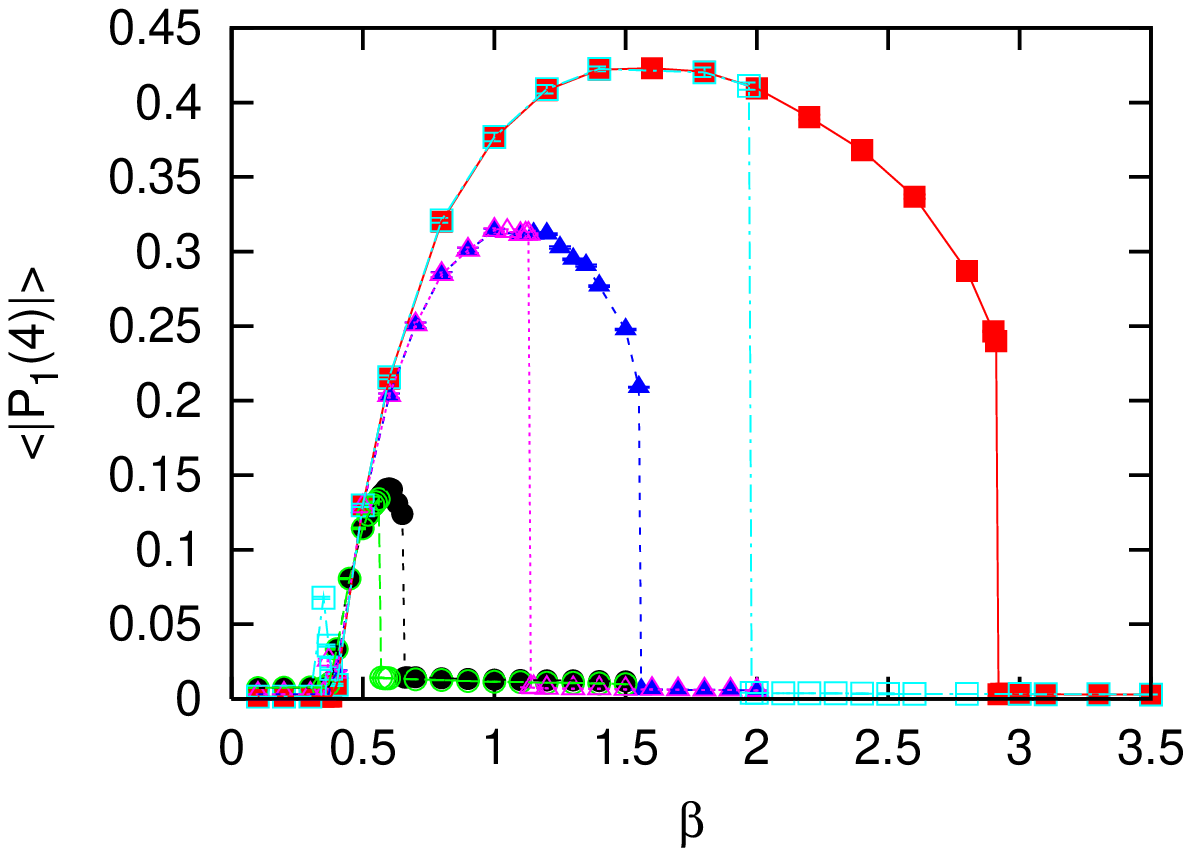,angle=270,width=7.4cm}
   \caption{The order parameter $\langle | P_1(n)| \rangle$
is plotted against $\beta$
for $n=2$ (left) and $n=4$ (right).
The system size is $N=15$ (circles), $N=25$ (triangles)
and $N=35$ (squares). The closed (open) symbols represent results
obtained with increasing (decreasing) $\beta$, which show
a clear hysteresis behavior.
}
  \label{polyakovline}
}

  \FIGURE{
\epsfig{file=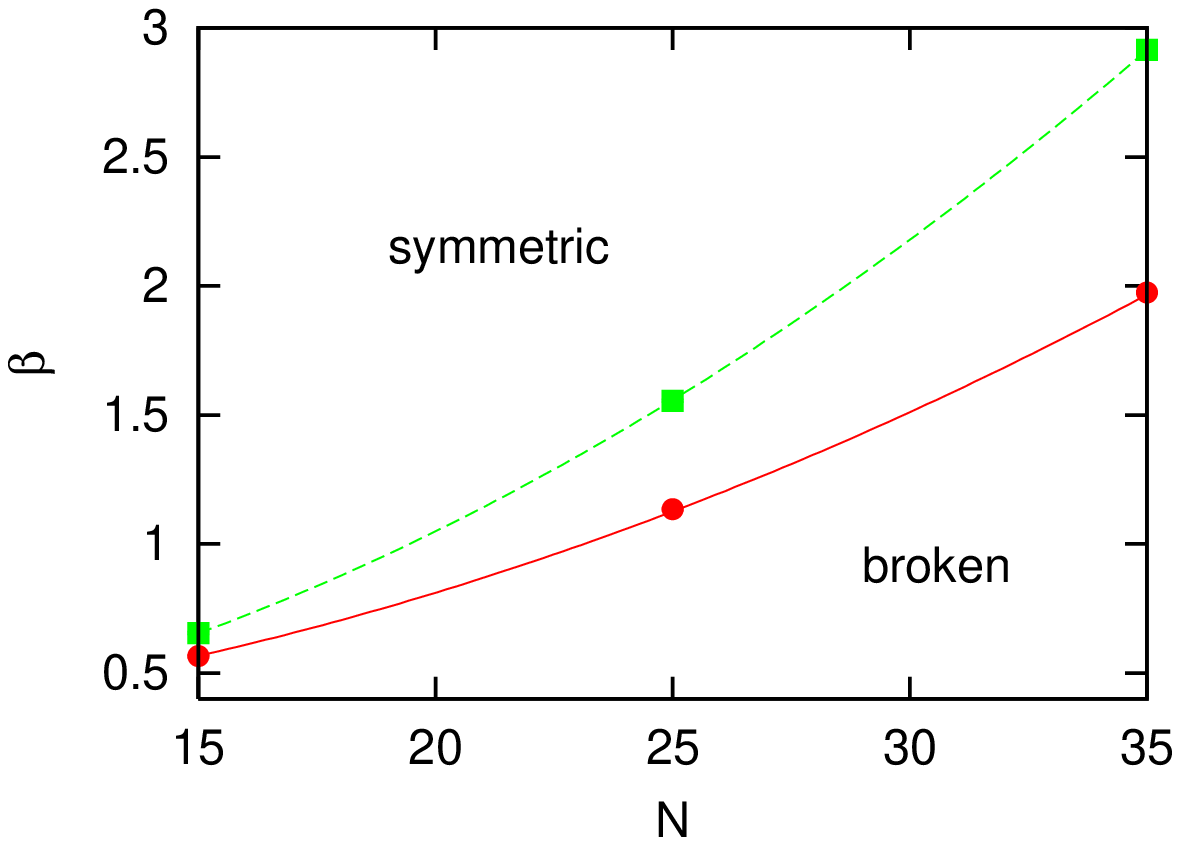,angle=270,width=7.4cm}
   \caption{
%
%
The upper (squares) and lower (circles) critical points
between the symmetric phase and the broken phase,
obtained from figure \ref{polyakovline}.
The lines re\-present
a fit to
$\beta = c_1 N^2 + c_2$, where 
$c_1 = 0.00226(1)$, $c_2 = 0.145(4)$ for the upper critical point, 
and $c_1 = 0.00141(1)$, $c_2 = 0.250(5)$ for the lower critical point.
}
  \label{beta_critical}
}

In figure \ref{polyakovline} we plot $\langle | P_1(n)| \rangle$
against $\beta$ for $n=2$ (left) and $n=4$ (right).
Since it turns out that $| P_1(n)|$ and $|P_2(n)|$ 
are not too different for each configuration\footnote{In principle,
it could also be that the U(1)$^2$ symmetry is broken down only to
U(1). In that case $| P_1(n)|$ and $|P_2(n)|$ would be very different
for each configuration, and taking the average over the two directions
would not be legitimate. We observed that this is not the case.},
we actually take an average over the two
NC directions to increase the statistics.
We observe that there is a phase, in which
the order parameter becomes non-zero.
On the other hand, the quantity $\langle | P_1(n)| \rangle$
for {\em odd} $n$ takes tiny values throughout the whole region of $\beta$.
This implies that the U(1)$^2$ symmetry is broken down 
to $({\rm Z}_2)^2$ in this phase, which we refer to as the ``broken phase''.
The fact that the U(1)$^2$ symmetry is unbroken,
both in the small $\beta$ regime
and the large $\beta$ regime,
can be understood from the strong coupling and 
the weak coupling expansions \cite{GAO}
as in the totally reduced model\footnote{A similar phase 
structure is 
obtained in the totally reduced model with the minimal 
twist \cite{ishikawa-okawa}.
In that case, however, $\langle | P_1(n)|\rangle$ 
for odd $n$ also acquires a non-zero value.
This is not so surprising since 
the argument based on the
momentum spectrum (\ref{momentum-polyakov}) 
does not apply to the model with the minimal twist.
}.

The transition between the strong coupling phase and
the broken phase occurs at $\beta \sim 0.35$ (for all $N$),
which coincides with the position of the gap in
figure \ref{fig:action} (left). Actually this value
agrees with the critical point of the 
bulk transition known in the 4d SU($N$) lattice gauge theory 
with Wilson's plaquette action at large $N$ \cite{Creutz:1981ip},
which is also reproduced by 
large-$N$ reduced models \cite{GAO,Nishimura:1996pe}.
Since the U(1)$^2$ symmetry is unbroken in our model 
in the strong coupling phase,
something must occur at the critical $\beta$ of the
bulk transition from the viewpoint of the Eguchi-Kawai equivalence.
What actually happens is
that the U(1)$^2$ symmetry is spontaneously broken
down to $({\rm Z}_2)^2$,
and the equivalence ceases to hold precisely at that point.
The transition between the broken phase and
the weak coupling phase is more important since
it is relevant for
the continuum limit. 
The hysteresis behavior, which
indicates a first order phase transition,
can now be seen clearly, unlike in figure \ref{fig:action}.
In fact the lower critical point of this phase transition,
denoted as $\beta_{\rm c}$ in the following,
can be estimated by perturbation theory,
and we obtain
\cite{KonagayaNishimura}
$\beta_{\rm c} \sim N^2$ at large $N$,
which is consistent with our results shown in figure \ref{beta_critical}.
This implies, in particular, that one always ends up in the
broken phase if one takes the $N\rightarrow \infty$ limit at fixed
$\beta$ (planar limit), which corresponds to $\theta =\infty$ in the 
context of NC gauge theory.


  \FIGURE{
\epsfig{file=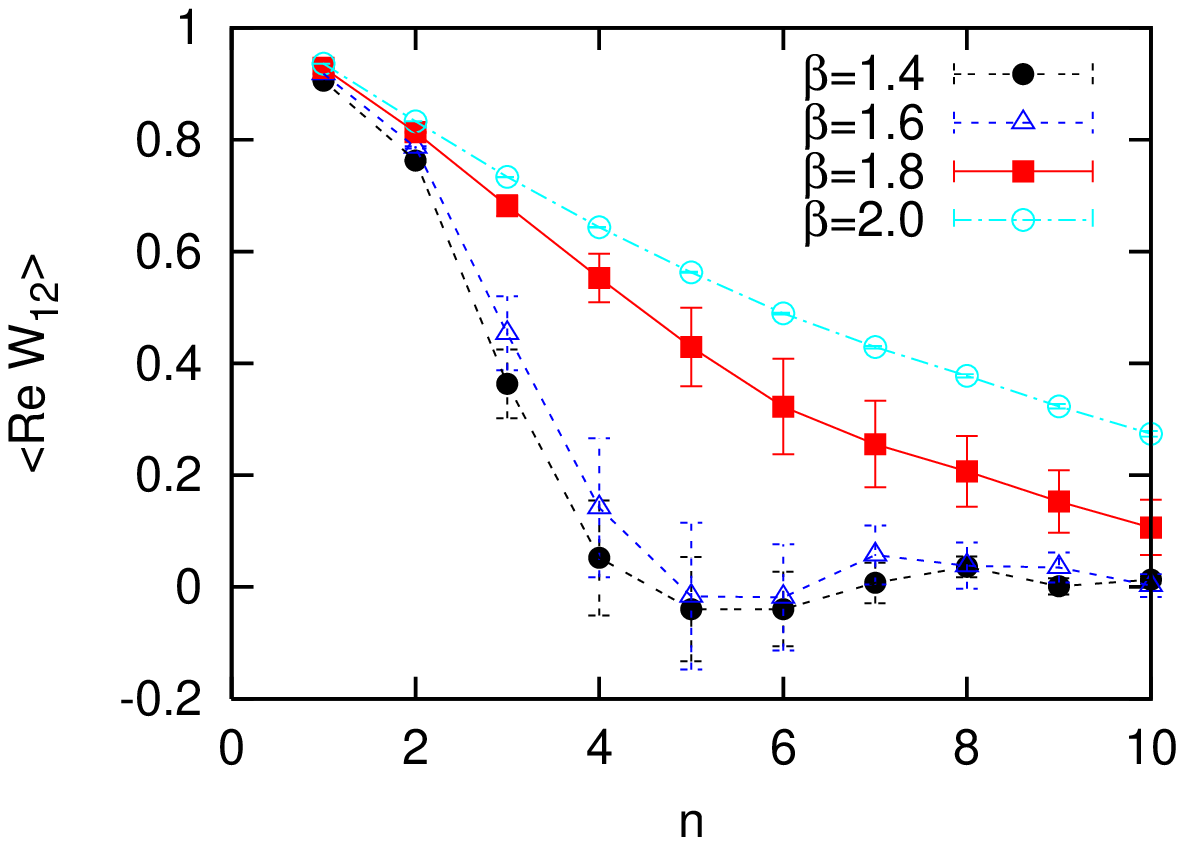,angle=270,width=7.4cm}
\epsfig{file=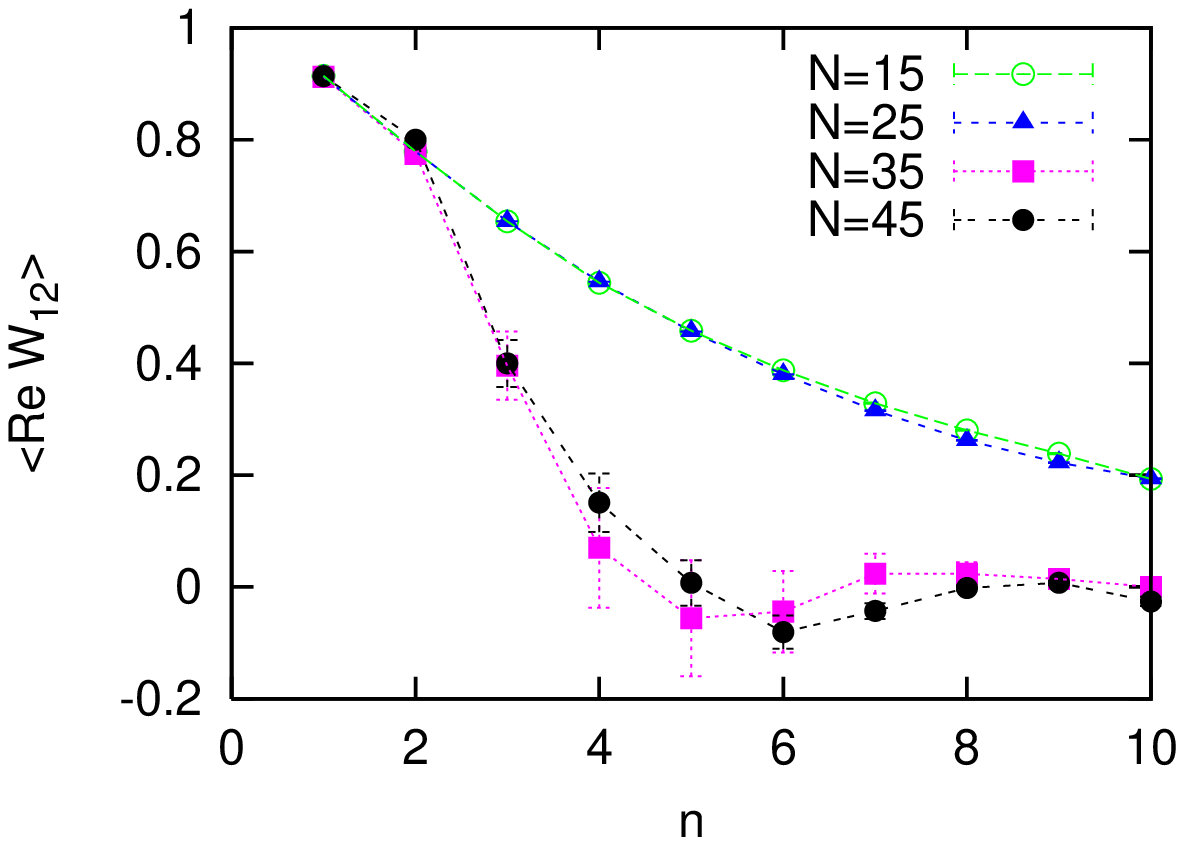,angle=270,width=7.4cm}
\epsfig{file=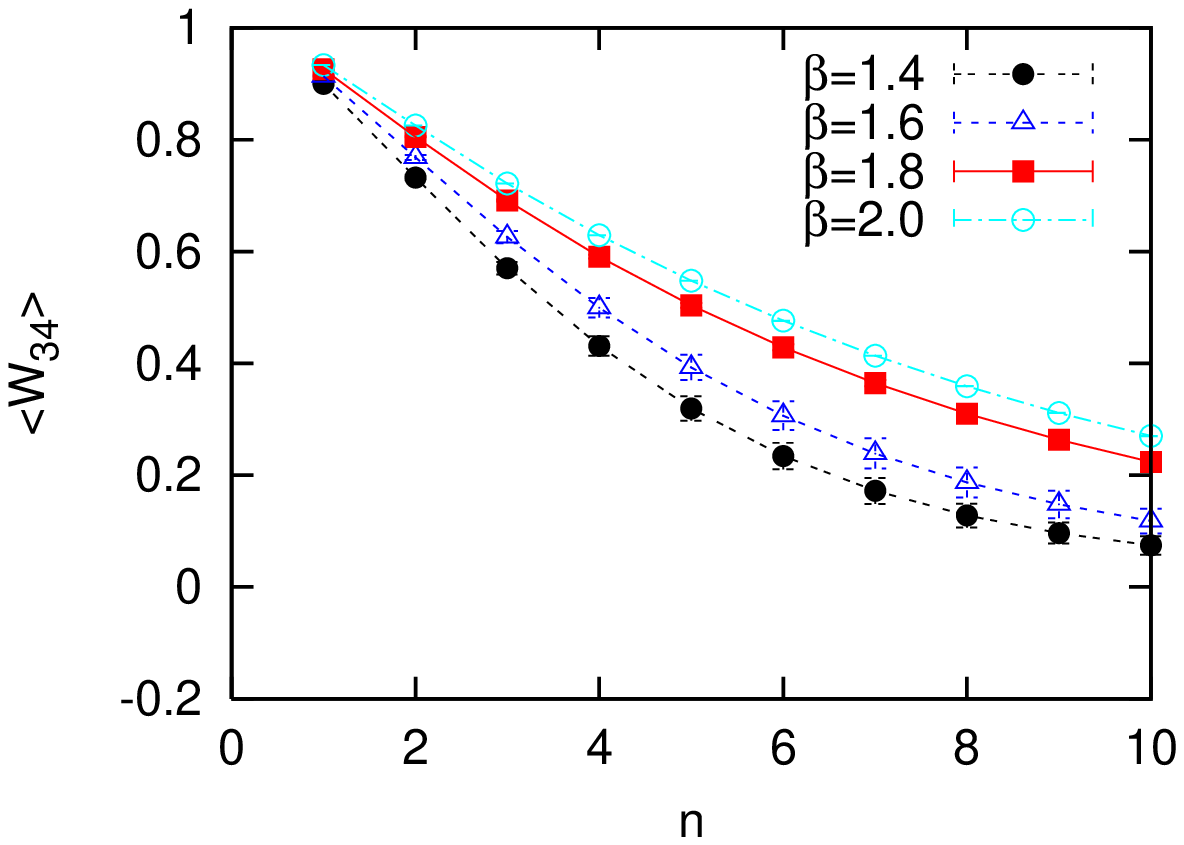,angle=270,width=7.4cm}
\epsfig{file=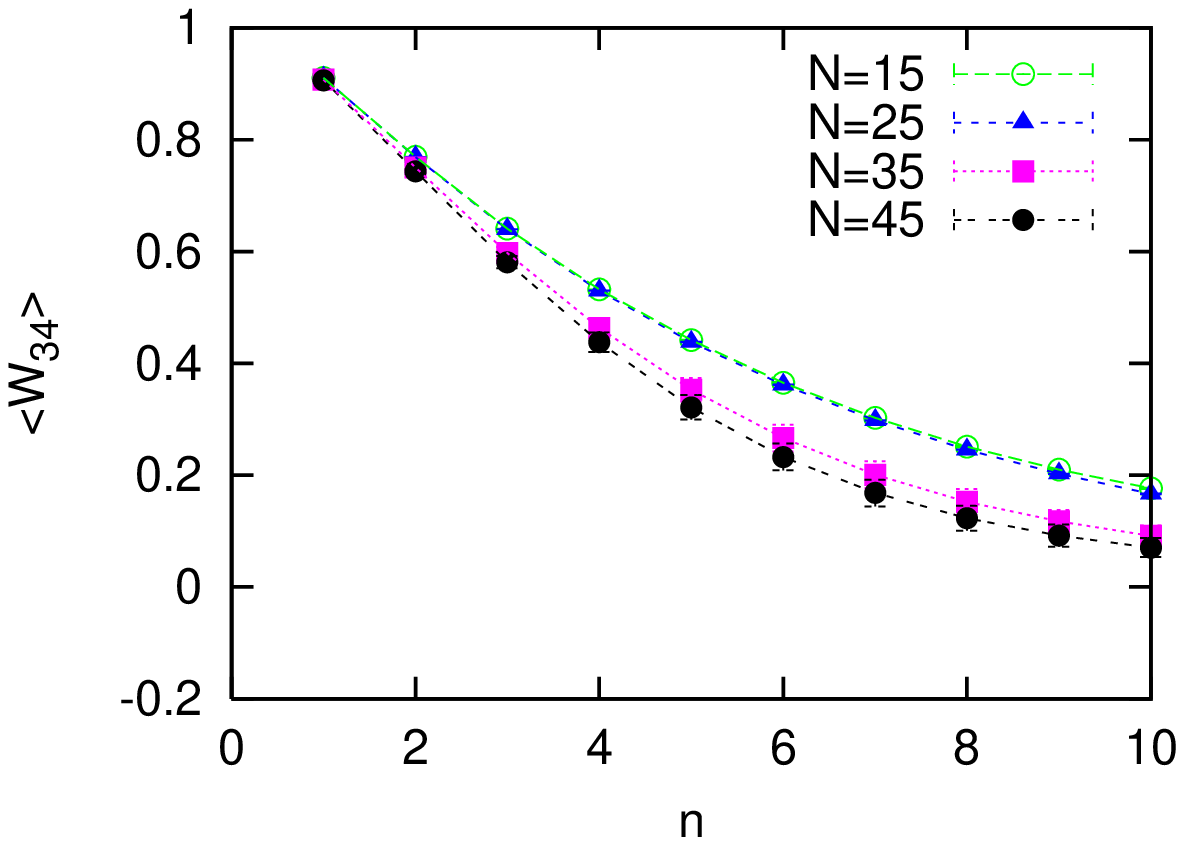,angle=270,width=7.4cm}
\epsfig{file=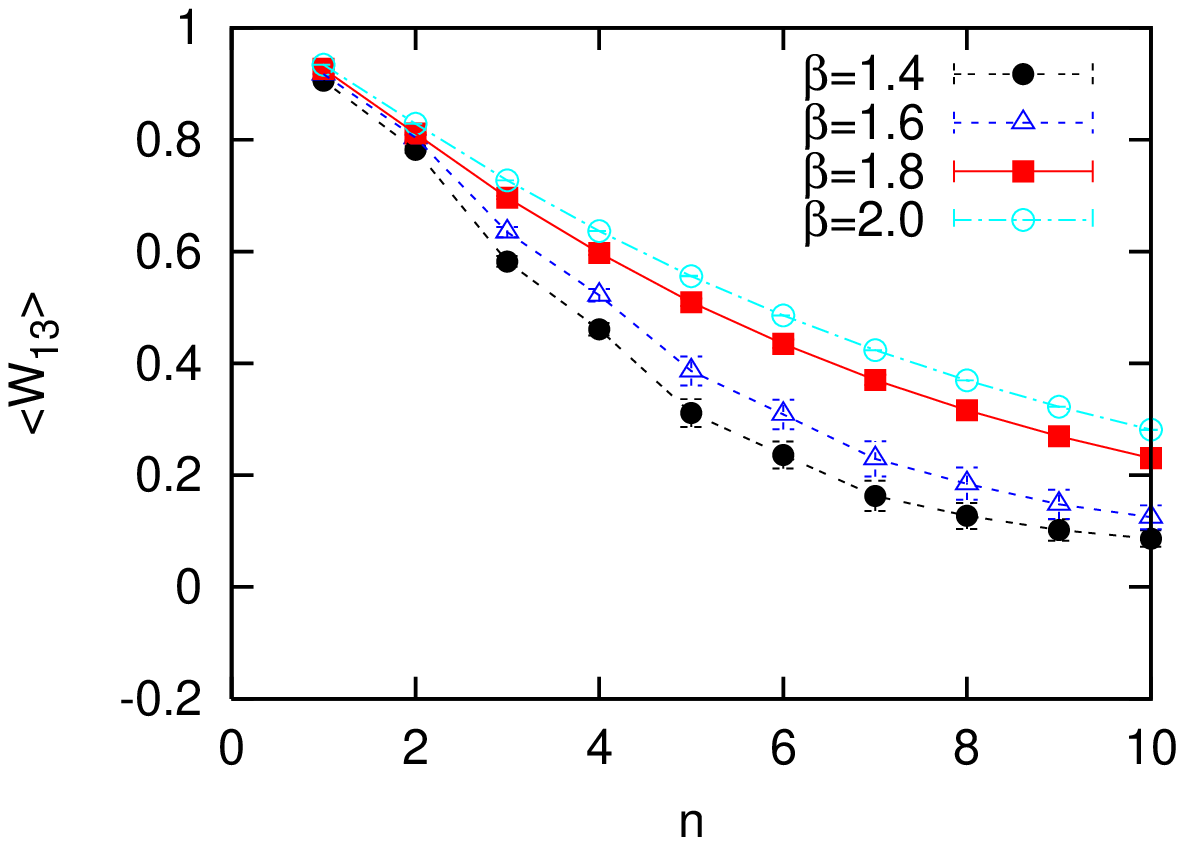,angle=270,width=7.4cm}
\epsfig{file=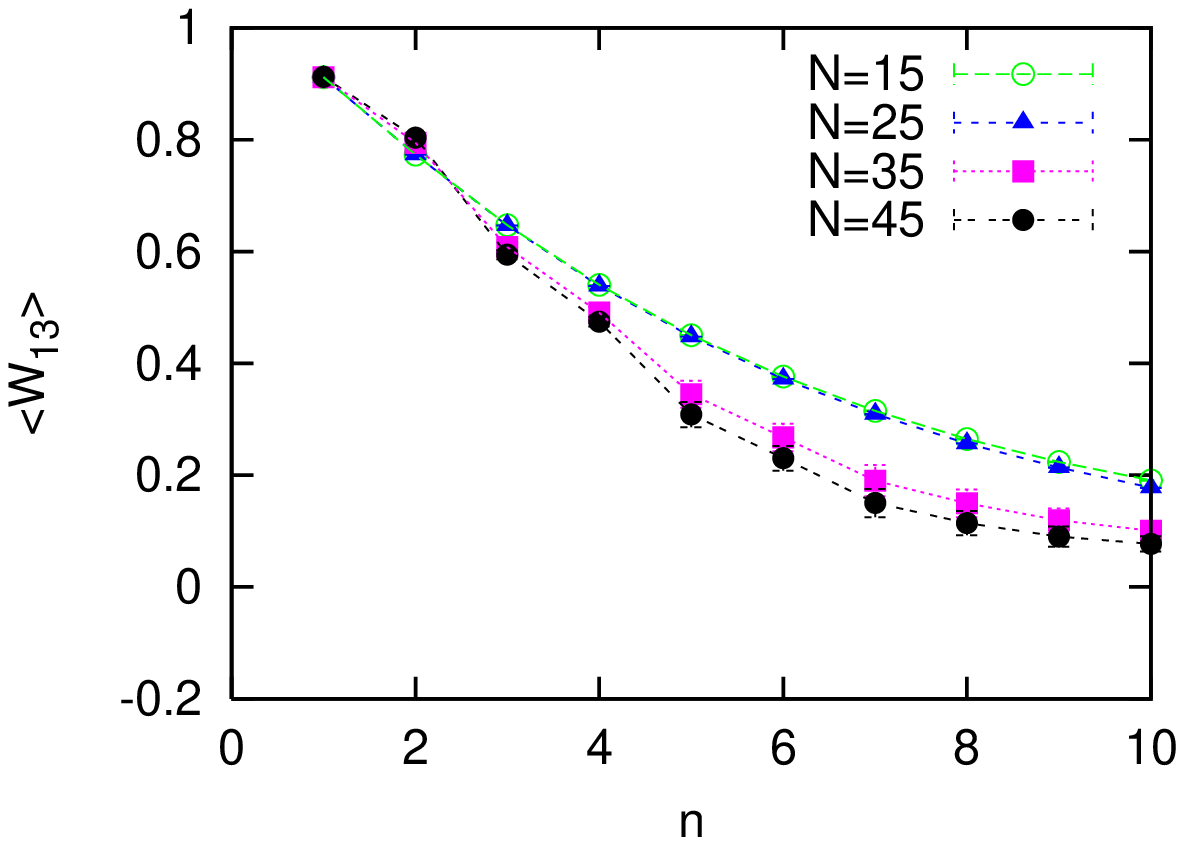,angle=270,width=7.4cm}
   \caption{(Left column) The expectation value of the Wilson loop
is plotted against its size $n$
for various $\beta$ at $N=35$.
The results for $\beta = 2.0$ are obtained
in the symmetric phase.
(Right column) The expectation value of the Wilson loop
is plotted against its size $n$
for $N=15,25,35,45$ at $\beta=1.5$.
The results for $N=15,25$ are obtained
in the symmetric phase.
}
  \label{wilson-beta}
}

Let us next consider closed Wilson loops, which play
an important role in commutative gauge theories as a criterion
for confinement.
In the present case, since we introduce non-commutativity
only in two directions,
there are three kinds of square-shaped Wilson loops depending on 
their orientations.
Using the parallel transporter
\beq
{\cal V}_\nu (z , n) \equiv
V_\nu (z) V_\nu (z + a \hat{\nu}) \cdots
V_\nu (z + (n-1) a \hat{\nu}) 
\eeq
in the commutative directions, the 
closed, square-shaped Wilson loops can be defined as
\beqa
W_{12}(n) 
&=&  ({\cal Z}_{12})^{n^2} \, \frac{1}{N L^2} \sum_z
\tr \Bigl( V_1 (z)^n \, V_2 (z)^n \, 
V_1 (z)^{\dag n} \, V_2 (z)^{\dag n} \Bigr)  \ ,  \nonumber \\
W_{\mu \nu}(n) 
&=&  \frac{1}{N L^2} \sum_z  
\tr \Bigl( V_\mu (z)^n \, 
{\cal V}_\nu (z , n) \, V_\mu (z + n a \hat{\nu})^{\dag n} \, 
{\cal V}_\nu (z , n)^\dag  \Bigr)  \ ,
\nonumber \\
W_{34}(n) 
&=& \frac{1}{N L^2} \sum_z  
\tr \Bigl( 
{\cal V}_3 (z , n) \, {\cal V}_4 (z+n a \hat{3}, n)
{\cal V}_3 (z+n a \hat{4},n)^\dag \, {\cal V}_4 (z, n)^\dag \Bigr) \ ,
\label{wilson-def}
\eeqa
where $\mu=1,2$ and $\nu=3,4$.
We may define $W_{\mu\nu}(n)$ for $\mu > \nu$ in a similar manner,
but it suffices to consider (\ref{wilson-def})
since $W_{\mu\nu}(n) = W_{\nu\mu}(n)^{*}$.
Mapping the expression (\ref{wilson-def}) back to the 4d NC space,
we find that 
the operator $W_{\mu\nu}(n)$ represents a closed $n \times n$ Wilson loop,
whose starting point is integrated
over the 4d space. 
Therefore it
does not carry non-zero momentum unlike the open Wilson line $P_\mu (n)$.
This corresponds to the fact that $W_{\mu\nu}(n)$ is invariant under
the U(1)$^2$ transformation (\ref{U1-2sym}), whereas $P_\mu (n)$ is not.

Let us consider the expectation value 
$\langle W_{\mu\nu}(n) \rangle$.
First we reduce the number of obser\-vables
using discrete symmetries.
Due to the symmetry under exchanging the third and the fourth 
directions, we have
$\langle W_{34}\rangle = \langle W_{43}\rangle \in \IR $ and
$\langle W_{\mu 3}\rangle = \langle W_{\mu 4}\rangle$,
where $\mu=1,2$.
Based on the symmetry under
$V_1(z) \mapsto V_2(z)$, $V_2(z) \mapsto V_1(z)^\dagger$,
we obtain
$
\langle W_{1 \nu}\rangle = \langle W_{2 \nu}\rangle \in \IR$,
where~$\nu=3,4$.
It then follows that
$\langle  W_{\mu\nu} \rangle$ with $\mu=1,2$ and $\nu=3,4$ are all
real and equal. We will use 
$\langle W_{13}\rangle$ as a representative, but
in actual simulation we measure
$\frac{1}{4} \sum_{\mu=1}^2 \sum_{\nu=3}^4 \langle  {\rm Re} \, 
W_{\mu\nu} \rangle$
to increase the statistics.
Note that $\langle W_{12}\rangle$, which
represents a closed Wilson loop  
in the NC directions, is complex in general
unlike in commutative gauge theories.
This occurs because the non-commutativity tensor $\Theta_{\mu\nu}$
breaks parity.

In figure \ref{wilson-beta} (left column) the 
expectation value of the Wilson loop 
is plotted against its linear size $n$
for various $\beta$ values at $N=35$.
First let us focus on the results at $\beta=2.0$, which are obtained 
in the symmetric phase. Here the three types of Wilson loop
are almost identical and follow the perimeter law
over the whole range of $n$ shown in the figure. 
This implies that the system is qualitatively similar 
to 4d compact U(1) lattice gauge theory in the commutative space, 
which is non-confining at weak coupling.
Note, however, that
the effect of the NC geometry is seen in the
dispersion relation for the same $\beta$ 
(figure \ref{eps:dispersion}).
As we decrease $\beta$, the system enters the
broken phase, and the three kinds of Wilson loops start to 
drift apart.

The right column of figure \ref{wilson-beta}
shows the results for various $N$ at $\beta=1.5$.
The data for $N=15$ and $N=25$,
which are obtained in the symmetric phase,
lie almost on top of each other.
This can be interpreted as the scaling behavior
corresponding to the would-be planar limit in the symmetric phase,
although 
one actually enters the broken phase as $N$ is increased further.
In fact the effect of increasing $N$ 
is similar to that of decreasing $\beta$,
which suggests the possibility to make the results
scale by increasing $\beta$ and $N$ simultaneously.
%

\section{Existence of a continuum limit}
\label{sec:double-scaling}

  \FIGURE{
\epsfig{file=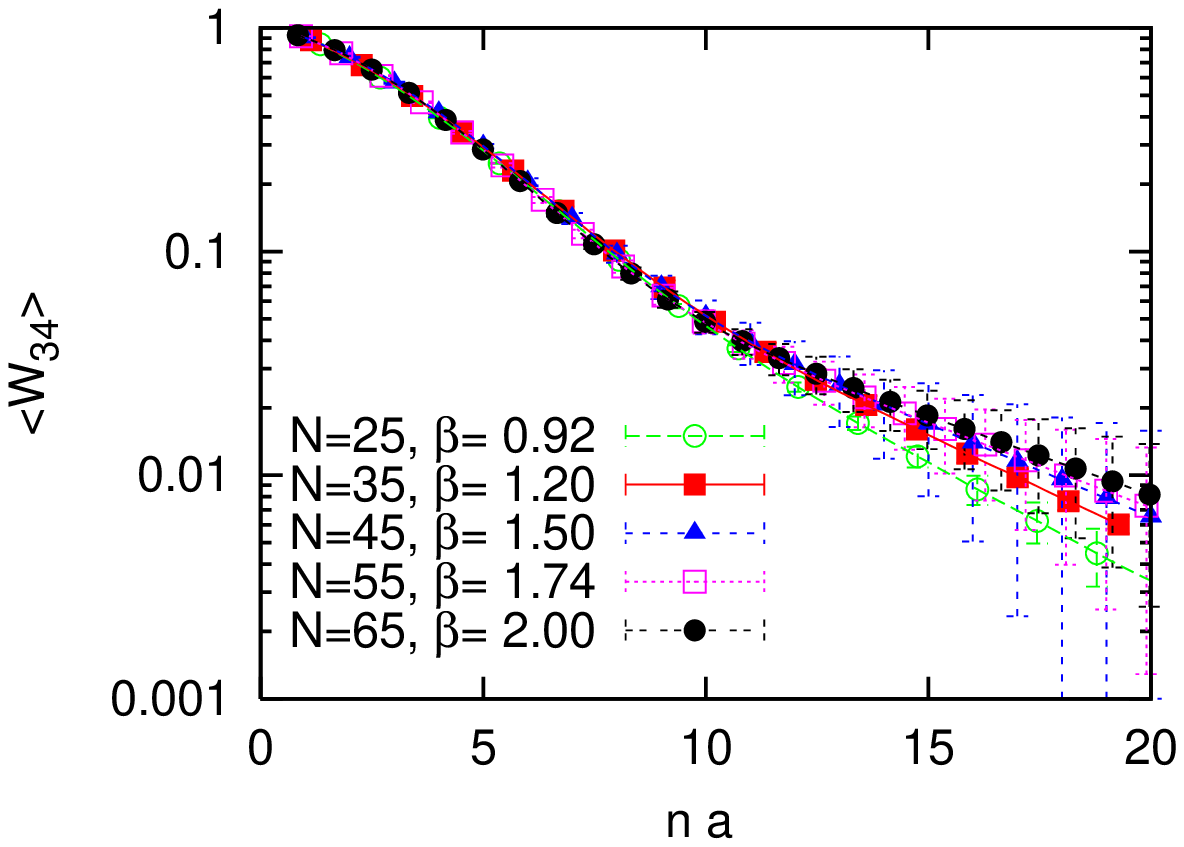,angle=270,width=7.4cm}
   \caption{The expectation value of the Wilson loop 
in the commutative plane. By tuning $\beta$
depending on $N$, we can make the results scale.
}
  \label{tameshi}
}

In this section we investigate whether it is possible
to tune $\beta$ as a function of $N$
in such a way that physical quantities scale.
For the rest of this paper
except for figure \ref{fig:extent-theta},
we set $a = 1$ for $N=45$ as a convention,
and the lattice spacing $a$ 
for other $N$ is determined\footnote{
If we 
had discarded the constraint (\ref{theta-def})
and determined $a$ for each $N$ to optimize the scaling,
we would have concluded that $\theta$ {\em had to be fixed}
in order to obtain the scaling behavior.
This follows from the fact that 
the scaling functions obtained for fixed $\theta$ 
have non-trivial $\theta$-dependence.
That $\theta$ does not receive renormalization
should therefore be considered as an observation
rather than an input of our study.
}
through (\ref{theta-def})
with $\theta=\frac{45}{\pi}\simeq 14.3$.


As a practical strategy to fine-tune $\beta$, we focus
on the expectation value of the square Wilson loop $W_{34}(n)$
in the commutative directions since it has a smooth dependence
on its size $n$.
Table \ref{beta-finetuning} shows the optimal values of $\beta$
for each $N$, and figure \ref{tameshi}
shows the corresponding plot.
The horizontal
axis represents the physical size ($na$) of the loop.
We observe a clear scaling behavior,
and the scaling region extends as $N$ increases.
In fact the optimal $\beta$ increases with $N$ much slower than
the lower critical point $\beta_c$ between the 
broken phase and the weak coupling phase, which grows as $N^2$
(see figure \ref{beta_critical}).
This implies that 
we remain in the broken phase in the double scaling limit.

\TABLE[pos]{\small%
\begin{tabular}{|c|c|c|c|}
\hline
\hline 
{~~~$N$~~~} 
&{~~~$a$~~~} & {~~~$\beta$~~~} \\ 
\hline 
\hline 
{25} 
& {1.34} & {0.92}  \\
\hline 
{35} 
& {1.13} & {1.20}  \\
\hline 
{45} 
& {1.00} & {1.50}  \\
 \hline
{55} 
& {0.90} & {1.74}  \\ 
\hline 
{65} 
& {0.83} & {2.00}  \\ 
\hline
\hline
\end{tabular}
\caption{The sets of parameters used 
for the double scaling limit.
\label{beta-finetuning}
}
}

Let us see whether other quantities
scale as well using the {\em same} sets of parameters
as those given in table \ref{beta-finetuning}.
In figure \ref{DSL_Wilsonloops_NC}
we plot the expectation value of the Wilson loop 
in the NC plane (left)
and in the mixed planes (right).
We do observe a compelling scaling behavior.

The Wilson loop in the commutative plane
follows the perimeter law at large size,
which suggests that the theory is non-confining in the commutative
directions.
The Wilson loop in the NC plane is complex in general,
and its real part oscillates around zero.
In figure \ref{DSL_Wilsonloops_NC-polar}
we plot the absolute value and the phase of the Wilson loop
against the physical area $A= (na)^2$.
The absolute value decreases monotonously obeying roughly 
the area law.
This is not so surprising since non-Abelian nature
comes in through the star product in (\ref{actNC}),
although we are dealing with a U(1) theory.
The phase $\Phi$ grows linearly as
$\Phi = \frac{1}{\theta} A$,
which is reminiscent of the Aharonov-Bohm effect
with the magnetic field $B=\theta^{-1}$ \cite{String}.
The same behavior has been observed in the 2d case \cite{2dU1}.

The behavior of the Wilson loop in the mixed directions 
is somehow between the other two kinds.
It decreases following roughly the perimeter law for large size,
but a slight oscillating behavior seems to be superimposed
(although these Wilson loops are real).
This suggests that the shape of the potential between
a quark and an anti-quark separated
in a NC direction is oscillating,
but we do not have a clear physical interpretation of such a behavior.

Let us turn to the open Wilson line,
which was studied in section \ref{sec:phase}
to investigate the spontaneous breakdown of U(1)$^2$ symmetry.
In figure \ref{DSLofPolyakov} (left) we plot 
the expectation value $\langle | P_1(n)|\rangle$ for even $n$
against the momentum defined by (\ref{momentum-polyakov}).
We observe a tendency that
the results lie on a single curve
except 
for the $n=2$ data (corresponding to the point 
at the smallest $p$ for each $N$).
A similar anomalous behavior is seen also 
in figure \ref{DSL_Wilsonloops_NC-polar} 
for $n = 3$ data.
We consider these behaviors to be finite $N$ artifacts
since they tend to disappear with increasing $N$.

  \FIGURE{
\epsfig{file=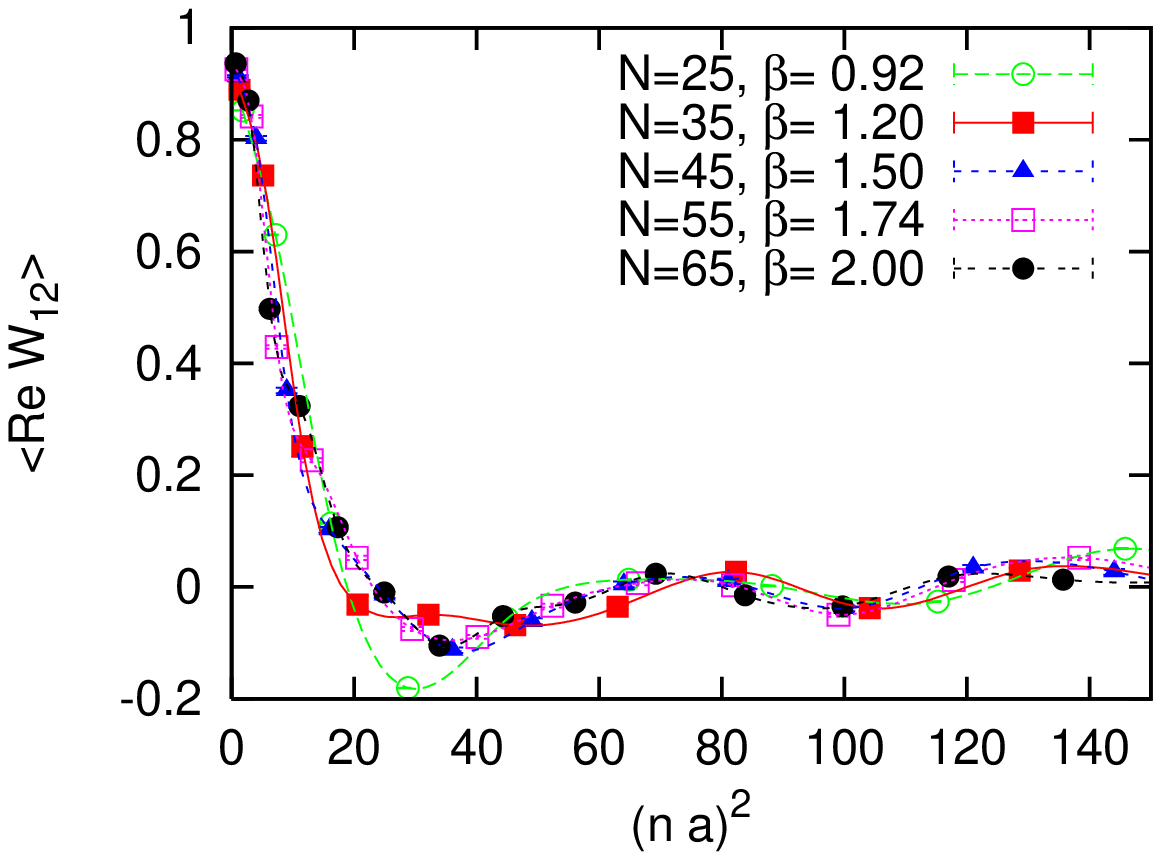,angle=270,width=7.4cm}
\epsfig{file=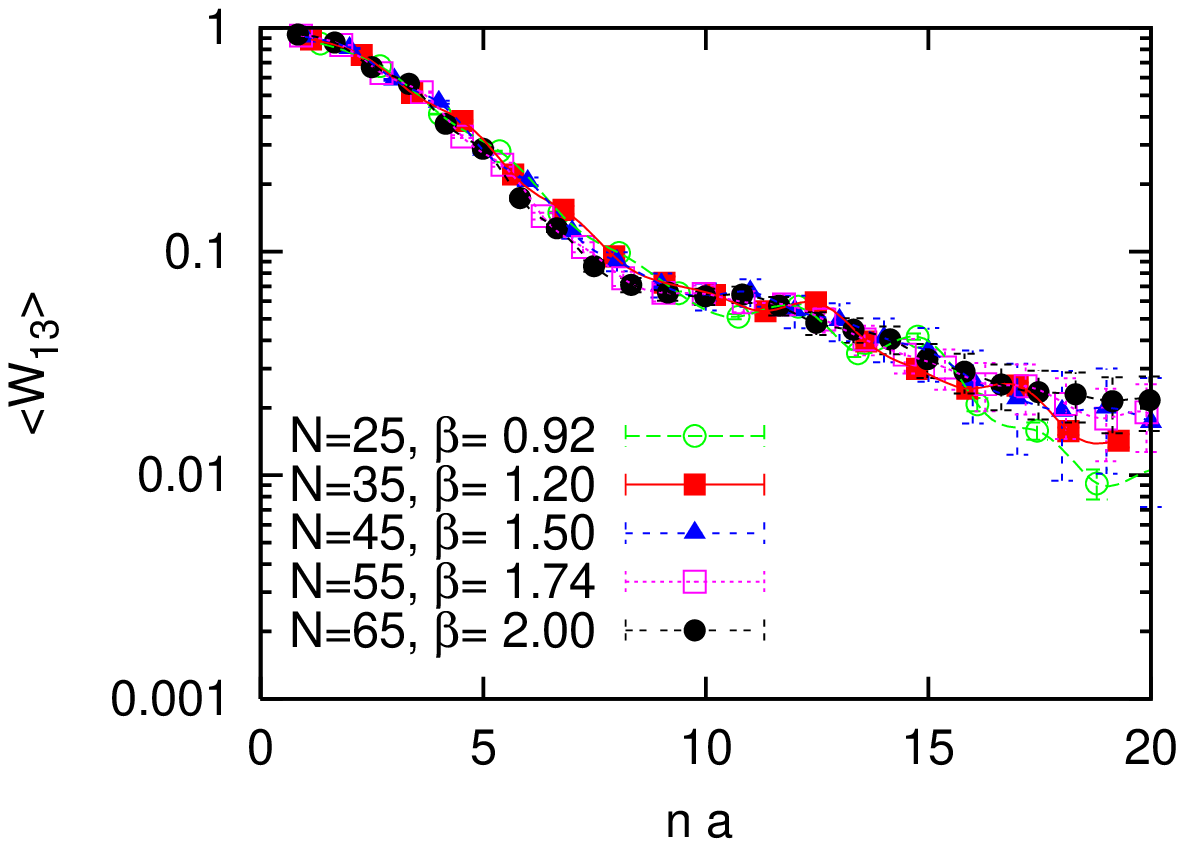,angle=270,width=7.4cm}
   \caption{The expectation value of the Wilson loop 
in the NC plane (left)
and in the mixed planes (right).
The values of $\beta$ are given in table \ref{beta-finetuning},
which was chosen in such a way that 
the expectation value of the Wilson loop
in the {\em commutative} plane scales.
}
  \label{DSL_Wilsonloops_NC}
}

  \FIGURE{
\epsfig{file=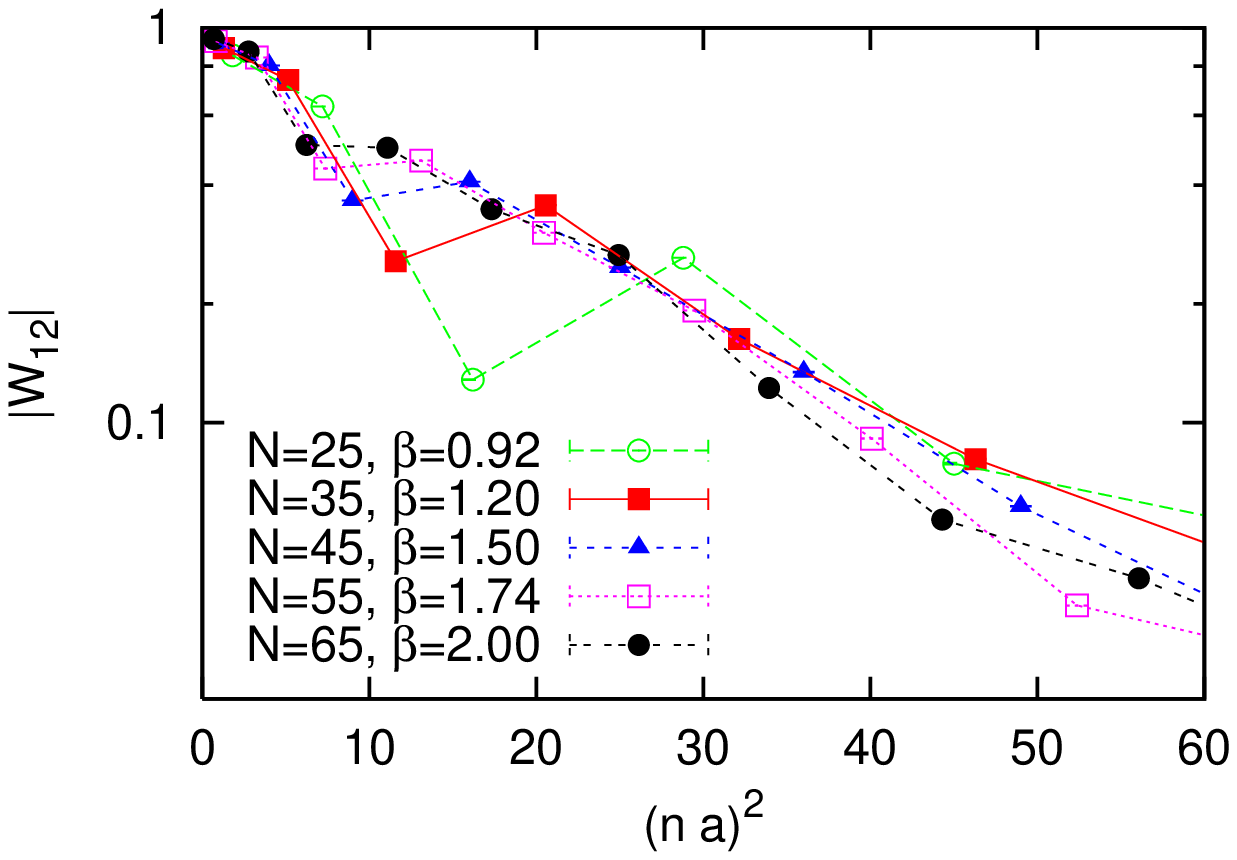,angle=270,width=7cm}
\epsfig{file=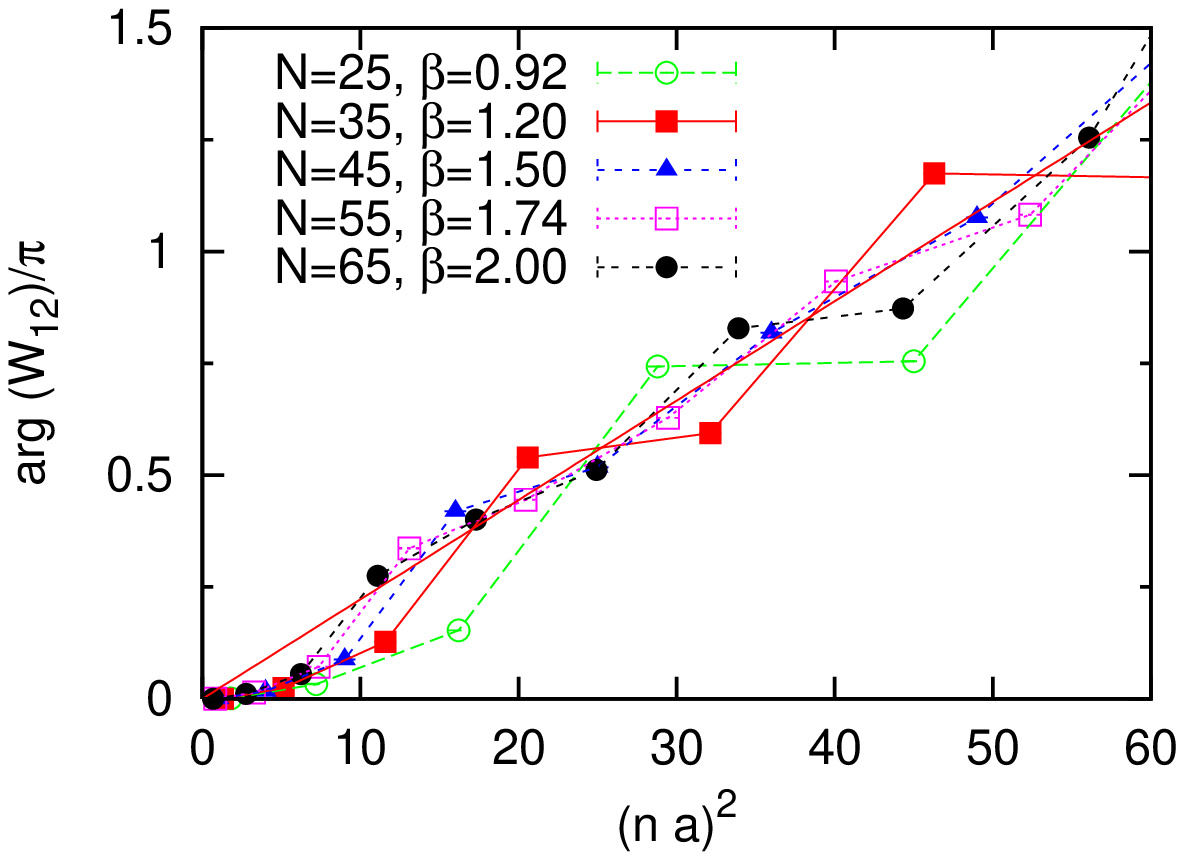,angle=270,width=7cm}
\caption{The absolute value (left) 
and the phase (right) of the Wilson loop
in the NC plane is plotted against
the physical area $A= (na)^2$.
The absolute value follows roughly an area law.
Beyond small areas,
the phase $\Phi$ agrees with the Aharonov-Bohm like behavior
$\Phi = A/\theta$ represented by the solid straight line.
}
  \label{DSL_Wilsonloops_NC-polar}
}

\section{Extent of the dynamical NC space}
\label{sec:phys-extent-NC}

Since the translational invariance 
in the NC directions is spontaneously broken in
the broken phase, it is natural to ask how the space
actually looks like in those directions.

As a related quantity, let us consider the eigenvalues
of the unitary matrix $V_\mu(z)$ ($\mu =1,2$),
which we denote 
as $e ^{i \vartheta_{\mu j}(z)}$ ($j = 1, \cdots , N$), 
where $- \pi < \vartheta_{\mu j}(z) \le \pi$.
The open Wilson line (\ref{orderparameter})
can be written in terms of the eigenvalues as
\beq
P_\mu(n) = 
\frac{1}{N L^2} \sum_z
\sum_{j=1}^N \ee^{i n \vartheta_{\mu j}(z)} \ .
\label{orderparameter2}
\eeq
Since the U(1)$^2$ transformation (\ref{U1-2sym}) rotates
all the eigenvalues
by a constant angle $\alpha_\mu$,
the eigenvalue distribution should be uniform 
if the U(1)$^2$ symmetry is not spontaneously broken.
This is how the open Wilson line
serves as an order parameter.
In the broken phase, we have seen that
the open Wilson line $P_\mu(n)$
acquires a non-zero expectation value only for even $n$.
This implies that the eigenvalues are clustered in two bunches
in a Z$_2$-symmetric way.
Since the translation in the NC direction is represented by
the U(1)$^2$ transformation, we may consider that the
eigenvalue distribution represents the ``shape'' of 
the NC space.

  \FIGURE{
\epsfig{file=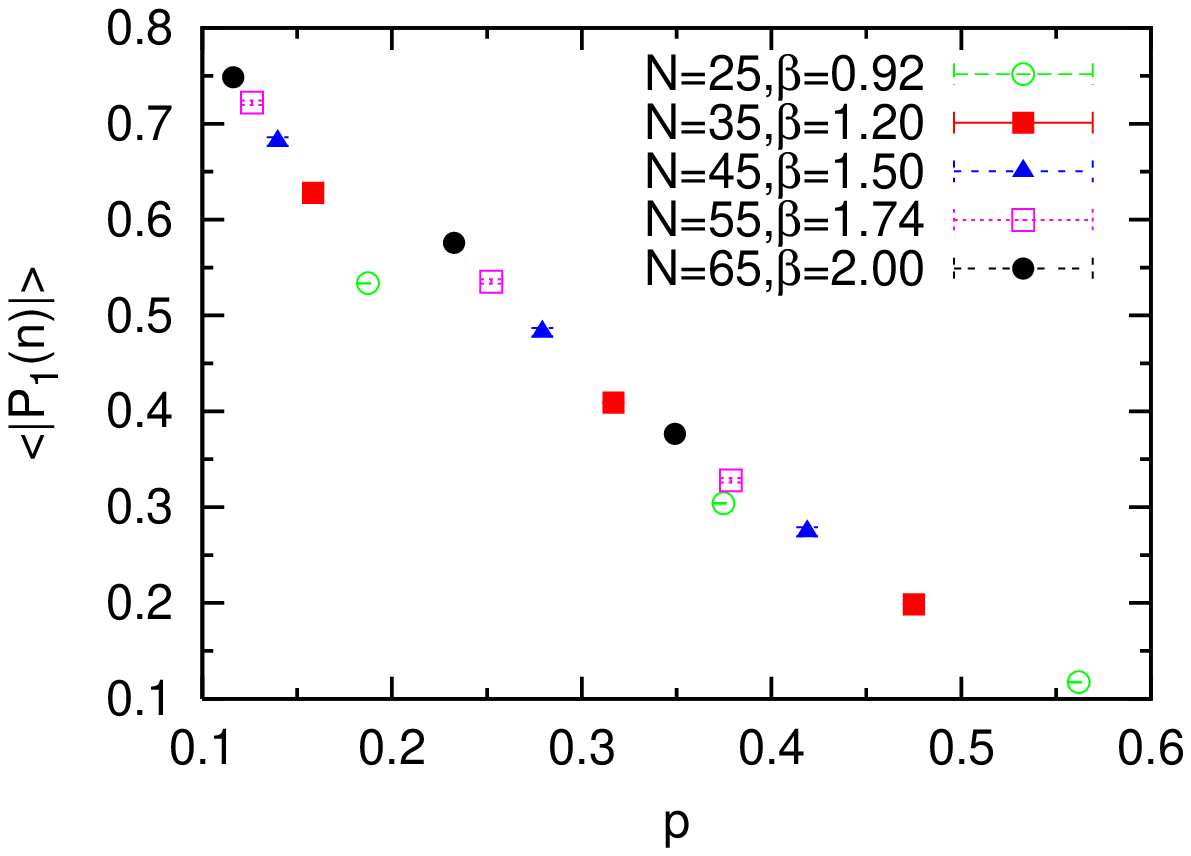,
angle=270,width=7.4cm}

\epsfig{file=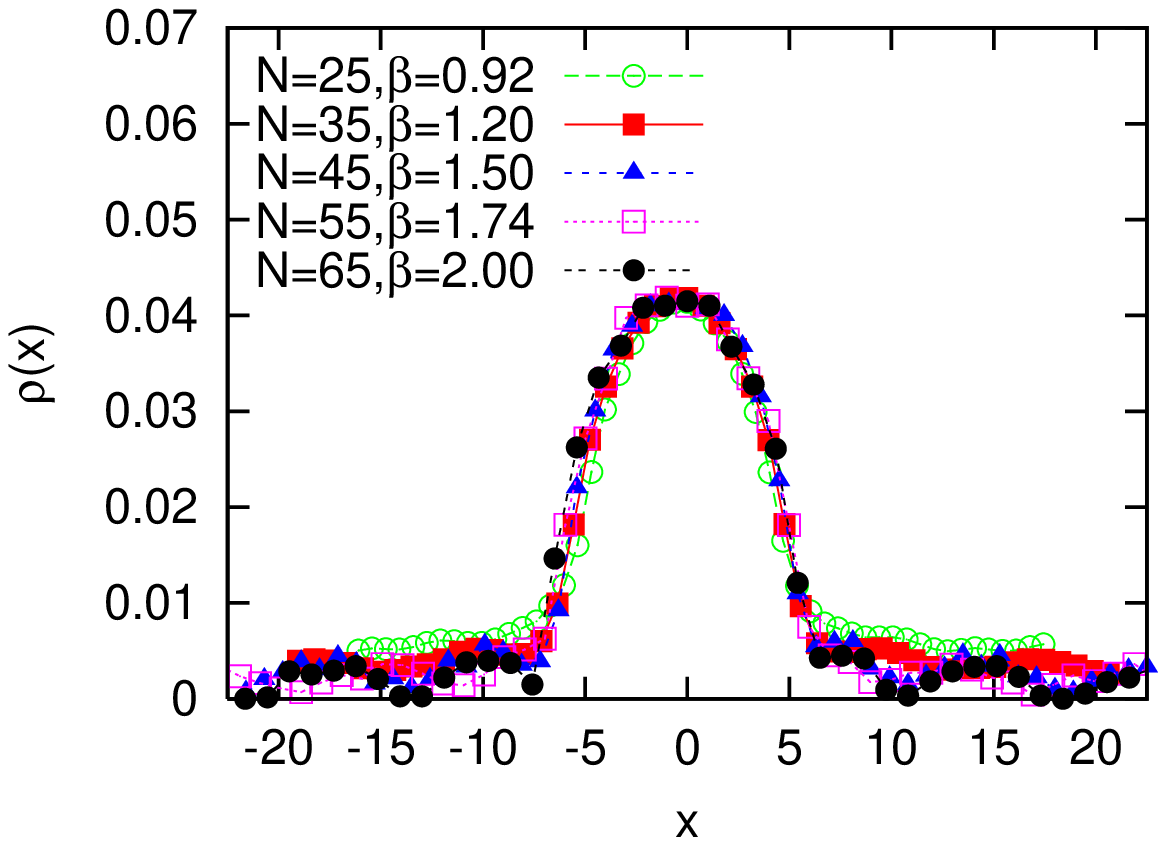,angle=270,width=7.4cm}
   \caption{(Left)
The expectation value $\langle | P_\mu(n)| \rangle$
for even $n$ is plotted 
against the momentum $p = \frac{\pi}{Na} n$
for the parameters listed
in table \ref{beta-finetuning}.
(Right) The eigenvalue distribution $\rho (x)$ 
is plotted 
for the parameters listed
in table \ref{beta-finetuning}.
%
}
  \label{DSLofPolyakov}
}

Let us define the eigenvalue distribution by
\beq
\rho (x) \equiv \frac{1}{2N L^2}  \sum_z \sum_{\mu=1}^2
\sum _{j=1}^N
\left\langle  \delta \left(x - \frac{Na}{\pi}
\vartheta_{\mu j}(z) \right) \right\rangle \ ,
\label{def-rho}
\eeq
where we have taken an average over the two NC directions
as we did for $\langle | P_1(n)| \rangle$.
The coefficient of $\vartheta_{\mu j} (z)$ is motivated from
the corresponding relation $p = \frac{\pi}{Na} n$ ($n$~:~even)
for the momentum conjugate to the coordinate $x$.
Before taking an ensemble average, we rotate each configuration
according to (\ref{U1-2sym}) in such a way that 
$\sum _z  \tr V_\mu (z)^2$ becomes 
real positive\footnote{We 
average over the Z$_2$-ambiguity in fixing the angle.}.
Since the eigenvalue distribution $\rho (x)$
is invariant under the shift $x \mapsto x+ Na$ modulo $2Na$,
we may restrict ourselves to the fundamental domain
$|x| \le Na/2$.
Figure \ref{DSLofPolyakov} (right) demonstrates
a clear scaling behavior of $\rho(x)$.
%
The eigenvalue distribution can be interpreted as the 
density distribution of D-branes \cite{rf:MVR}.
If we take the view point of string theory, in which the
space-time is represented by the eigenvalue distribution of the
matrices \cite{SSB,related},
our results imply that the ``dynamical space'' in the NC directions has
shrunk, but it has a finite extent in the double scaling limit.

Since we have seen that the extent of the ``dynamical space''
in the NC direction is finite for a fixed $\theta$,
it is natural to ask how the extent depends on $\theta$.
Let us note that
\beqa
\frac{1}{2}\sum_{\mu=1}^2 \langle|P_\mu(2)|\rangle 
&=& 2 \int _{-Na/2}^{Na/2} dx \, e^{\frac{2 \pi i }{Na}x } \rho(x) 
\nonumber \\
&=& 1 - \left(\frac{2\pi}{Na}\right)^2
 \int _{-Na/2}^{Na/2} dx \, x^2 \rho(x) + \cdots \ , 
\eeqa
where taking the absolute value on the left-hand side
corresponds to rotating $\vartheta_{\mu j}(z)$ in the
definition (\ref{def-rho}) of $\rho(x)$ 
before taking the ensemble average as explained above.
In the last line, we have assumed that $\rho(x)$ is 
peaked around $x=0$ in the fundamental domain $|x| \le Na/2$.
  \FIGURE{
\epsfig{file=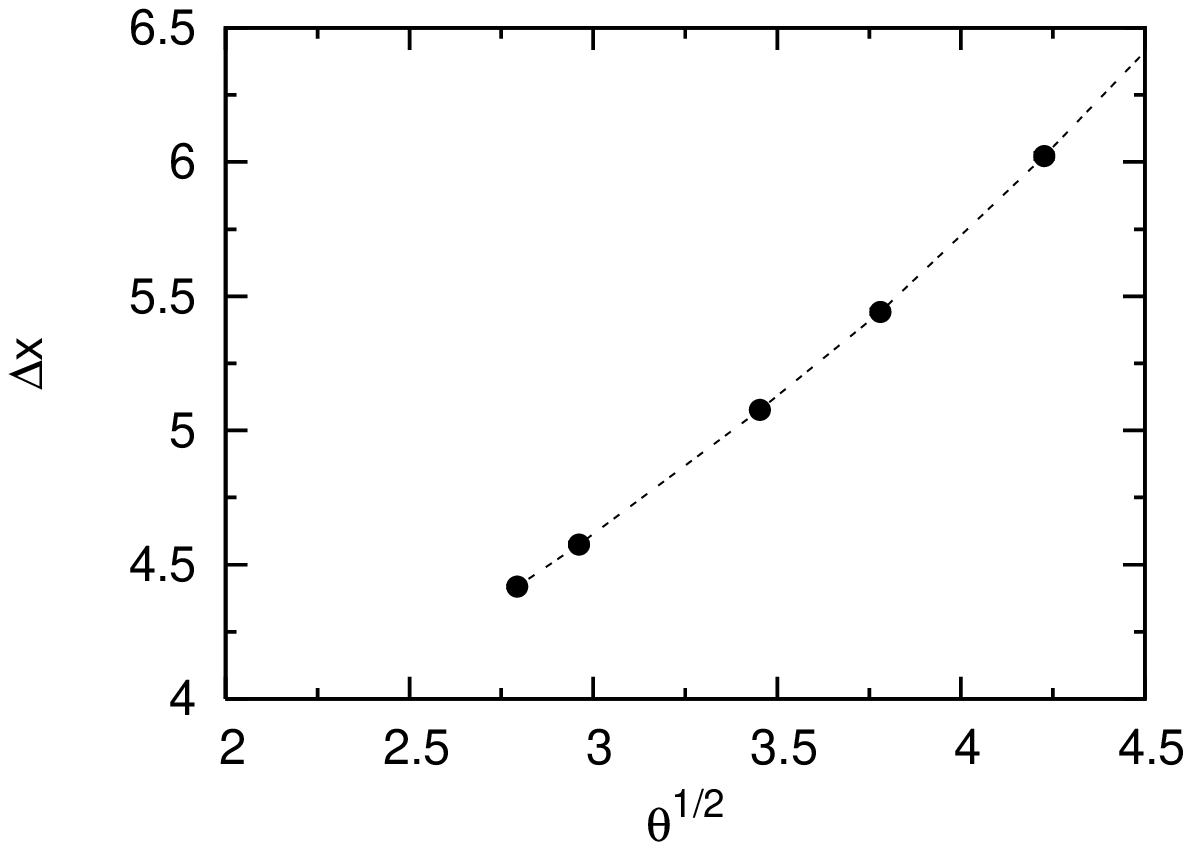,angle=270,width=7.4cm}
   \caption{
The extent of the dynamical space 
in the NC directions defined by
(\ref{NCextent}) from the view point of string theory,
is plotted against $\sqrt{\theta}$ 
for $N=35$.
}
  \label{fig:extent-theta}
}
As a definition for
the extent of the dynamical space in the NC direction, 
we therefore use
\beq
\Delta x \equiv \frac{Na}{\pi}
\sqrt{\frac{1}{2}
\left ( 1-  
\frac{1}{2}\sum_{\mu=1}^2
\langle|P_\mu(2)|\rangle \right)}   \ .
\label{NCextent}
\eeq
It should be mentioned that the space-time, on which the NC gauge theory
is defined, has the extent $Na$, which diverges in the double scaling limit.
However, since the translational invariance in the NC directions
is spontaneously broken, the observer on the NC space-time does not
recognize that the space-time extends to infinity.
The quantity $\Delta x$ measures the extent of the space-time
in the NC directions,
which is recognized by the observer to be qualitatively uniform.
In that sense, $\Delta x$ is analogous to the width of the stripes
\cite{Bietenholz:2004xs} in the 3d NC $\lambda \phi^4$ theory,
in which the translational invariance in the NC directions
is broken due to the space-dependent vacuum expectation value
of the scalar field.

Instead of repeating the whole procedure of
taking the double scaling limit at each $\theta$,
here we assume that the double scaling
is obtained for the same sets\footnote{In fact
one can always multiply the lattice spacing $a$
by a $\theta$-dependent factor without affecting the
scaling property. This ambiguity corresponds to
the arbitrary choice of the $\Lambda$-parameter 
at each $\theta$. The assumption we adopted here
corresponds to taking the $\Lambda$-parameter
independent of $\theta$.
} 
of $(a,\beta)$ listed
in table \ref{beta-finetuning}.
This assumption is justified at large $N$ 
given that the ultraviolet properties of NC theories
are independent of $\theta$.
For various values of $\beta$ in the range $0.92 \le \beta \le 2.00$,
we determine $a$ by
interpolating the relation between $a$ and $\beta$ 
presented in table \ref{beta-finetuning}.
The value of $\theta$ is then determined from (\ref{theta-def}) 
using $a$ and $N$.
In figure \ref{fig:extent-theta} we plot $\Delta x$ 
against $\sqrt{\theta}$ for $N=35$.
At large $\theta$ the extent $\Delta x$ increases linearly 
with $\sqrt{\theta}$ as expected on dimensional account.

\section{Dispersion relation}
\label{sec:dispersion-rel}

In this section we investigate the dispersion relation
in the symmetric phase and the broken phase separately.
The ana\-lysis in the symmetric phase reveals
the IR singularity (\ref{ea}), which is 
responsible for the phase transition,
whereas the analysis in the broken phase
enables us to identify the Nambu-Goldstone mode
associated with the spontaneous breakdown of the
U(1)$^2$ symmetry.
Thanks to the existence of the commutative directions in the present
setup, we may regard one of the coordinates (say, $x_4$) 
as ``time''.
From the exponential decay of the two-point correlation function 
of open Wilson line operators separated in the ``time'' direction,
we can extract the energy of a state that couples to the operator.
Similar studies have been done also
in the case of NC scalar field theory \cite{Bietenholz:2004xs}.

  \FIGURE{
\epsfig{file=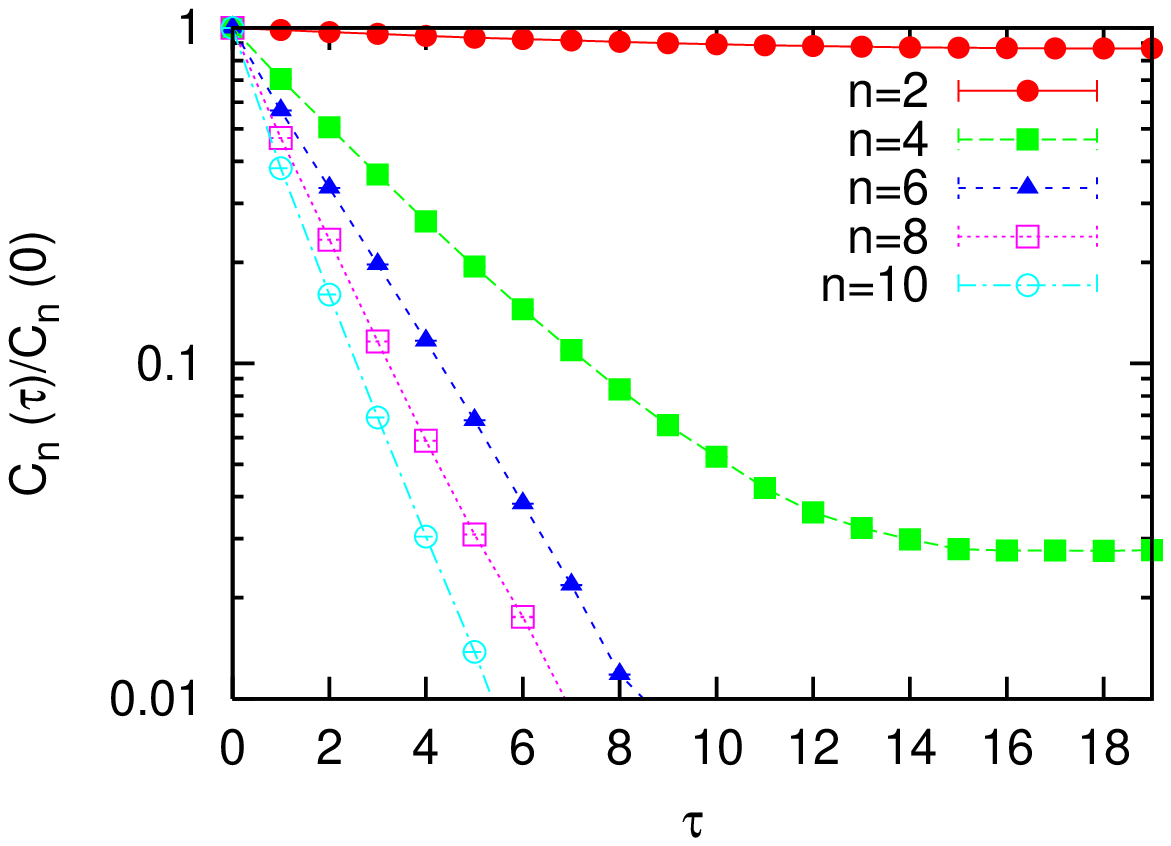,angle=270,width=7.4cm}
\epsfig{file=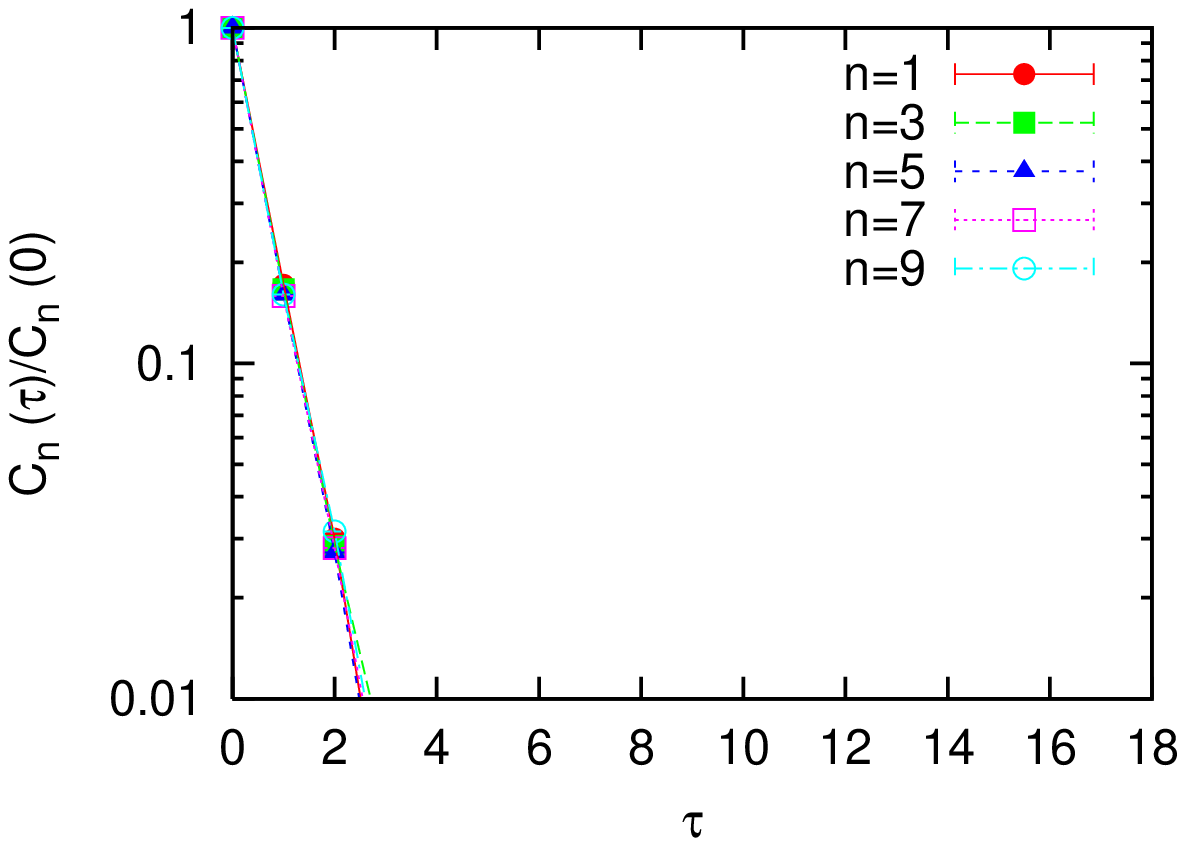,angle=270,width=7.4cm}
   \caption{Two-point correlation function of the open Wilson lines
for even $n$ (left) and odd $n$ (right) is plotted against
the separation $\tau$ for $N=35$ and $\beta = 2.00$. 
The vertical axis is normalized in such a way
that the function starts off from unity at the origin.
}
  \label{eps:correlationPP}
}

\subsection{Results in the symmetric phase --- IR singularity}

  \FIGURE{
\epsfig{file=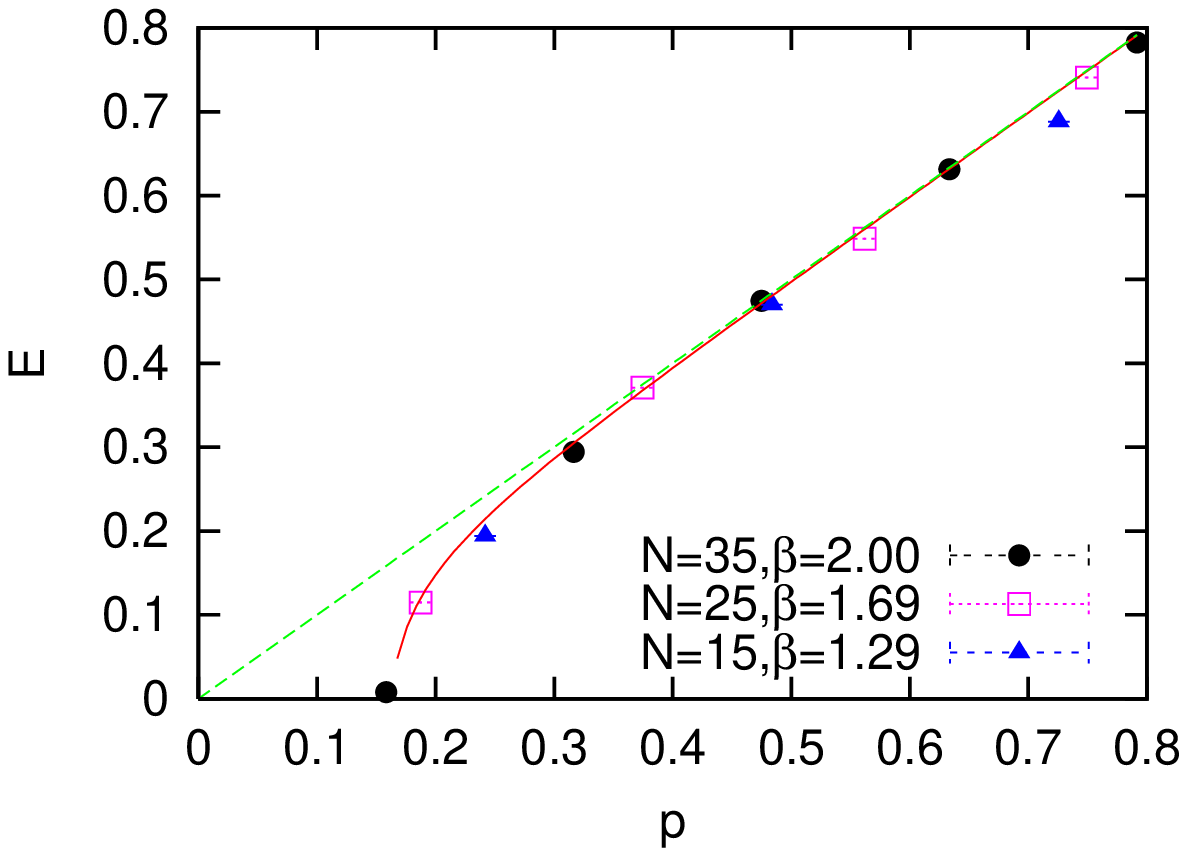,
angle=270,width=7.4cm}
   \caption{The dispersion relation in the symmetric phase.
The energy $E$ obtained from the 
two-point correlation function (\ref{2-pt-cor})
is plotted against the momentum $p$
for $(N,a,\beta)=(35,1.13,2.00)$, $(25,1.32,1.69)$,
$(15,1.73,1.29)$.
%
%
}
  \label{eps:dispersion}
}

Let us define 
the open Wilson line operator at a fixed time $x_4$ as
\beq
P_\mu(x_4 , n) \equiv  \frac{1}{NL} 
\sum_{x_3} \tr \Bigl( V_\mu(x_3,x_4) ^n \Bigr) 
 ~~~\mbox{for}~ \mu=1,2 \ ,
\label{openWilsonlinesdef}
\eeq
which has a zero momentum component 
in the $x_3$ direction\footnote{One can 
introduce a non-zero momentum component $p_3$
in the $x_3$ direction
by inserting a phase factor $\ee^{i p_3 x_3}$ in the summation
over $x_3$ in eq.\ (\ref{openWilsonlinesdef}).
The dispersion relation (\ref{NCdispertionat1loop}) will then
have a term $(p_3)^2$ on the right-hand side.}
and a non-zero momentum component 
(\ref{momentum-polyakov}) in a NC direction
depending on $n$.
Then we define the two-point
correlation function of the open Wilson lines
\beq
C_n (\tau) \equiv \frac{1}{2} \sum_{\mu=1}^2 \sum_{x_4}
\Bigl\langle P_\mu(x_4 , n)^{*} \cdot 
 P_\mu(x_4 + \tau , n)  \Bigr\rangle
\label{2-pt-cor}
\eeq
with a separation $\tau$ in the temporal direction.
In actual measurement we also consider the case with the 
roles of $x_3$ and $x_4$ exchanged, and take an average over the two
cases to increase the statistics.

In figure \ref{eps:correlationPP}
we plot the two-point correlation function
for even $n$ (left) and odd $n$ (right).
For even $n$ we observe clear exponential behaviors
$\ee ^{- \lambda \tau}$, from which we extract the energy 
$E = \lambda / a$ at each momentum $p = \frac{\pi}{Na} n$.
For odd $n$ the decay is very rapid, which suggests
that the energy (as well as the momentum) is on the
cutoff scale.

In figure \ref{eps:dispersion}
we show the dispersion relation 
obtained from the 
two-point correlation function (\ref{2-pt-cor}) for even $n$.
The set of parameters $(N,a,\beta)$ is chosen
as in the broken phase.
Namely, $a$ is determined through (\ref{theta-def})
with $\theta=\frac{45}{\pi}\simeq 14.3$, and 
we fine-tune $\beta$ at each $N$ in such a way that
the scaling behavior of the Wilson loops 
in the commutative plane is optimized.
It turns out that the data points $(E,p)$ for different $N$
lie to a good approximation on a single curve
\beq
E^2= p^2 - \frac{c}{(\theta p)^2}
\label{NCdispertionat1loop}
\eeq
with $c\simeq 0.1285$.
This form of the dispersion relation
is expected from the one-loop calculation of 
the vacuum polarization \cite{mst,LLT,ruiz,Bassetto:2001vf}.
Due to the negative sign
of the second term in (\ref{NCdispertionat1loop}),
the usual Lorentz invariant (massless)
dispersion relation is bent down.
The IR singularity is regularized 
on the finite lattice
since the smallest non-zero momentum 
(which corresponds to $n=2$) is given
by $\frac{2\pi}{Na}\propto \frac{1}{\sqrt{N}}$.
However, if one attempts to increase $N$ further,
the energy at the smallest non-zero momentum vanishes at some $N$,
and one enters the broken phase.
Therefore we cannot take the double scaling limit in 
the symmetric phase.
The scaling observed in the symmetric phase represents an effective
theory with a finite cutoff.
In the case of NC scalar field theory \cite{Bietenholz:2004xs},
the double scaling limit can be taken in the symmetric phase
since the IR singularity appears in the dispersion relation
with the positive sign.

\subsection{Results in the broken phase --- Nambu-Goldstone mode}

  \FIGURE{
\epsfig{file=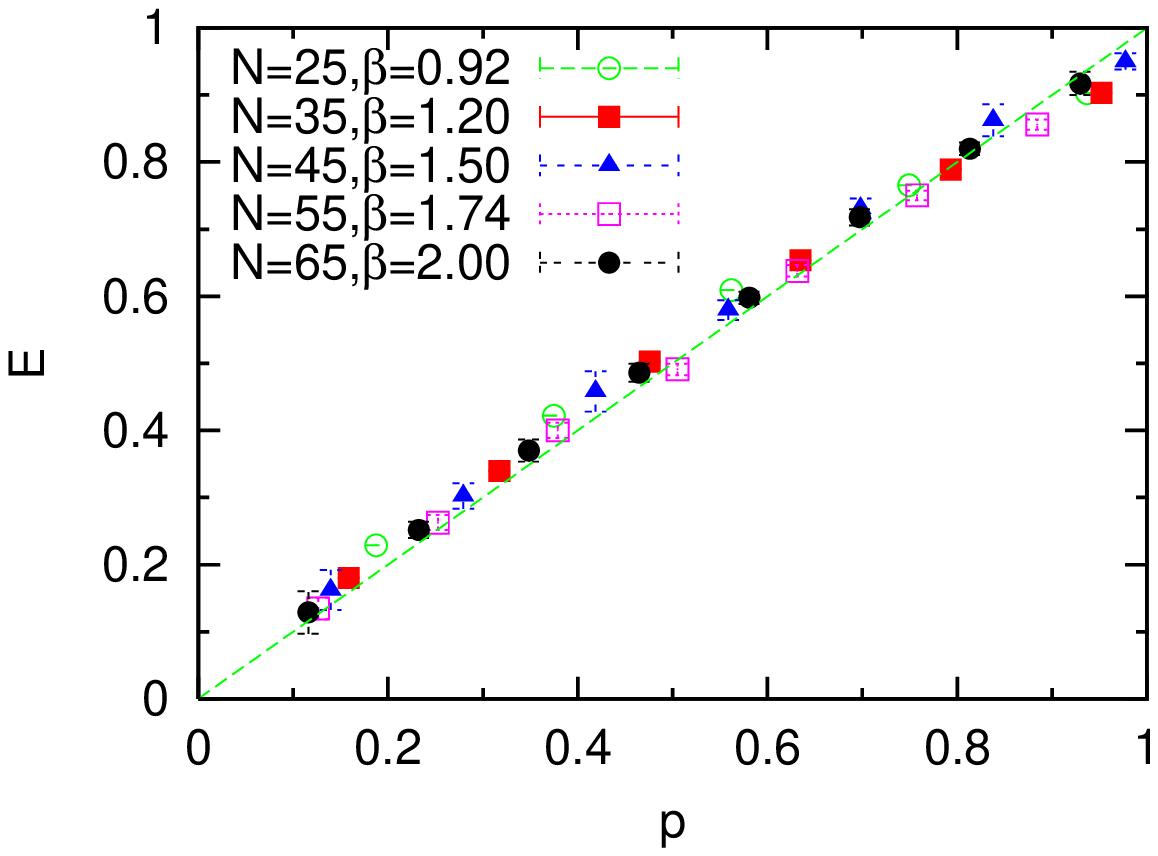,
angle=270,width=7.4cm}
   \caption{The dispersion relation in the broken phase.
The energy $E$ obtained from the 
two-point correlation function (\ref{2-pt-cor})
is plotted against the momentum $p$ for the parameters listed
in table \ref{beta-finetuning}.
}
  \label{eps:dispersion-broken}
}

When we extend the study of the dispersion relation
to the broken phase, we should note that 
the momentum components in the NC directions are no longer conserved,
although the momentum component
in the {\em commutative} direction still is.

Let us therefore define 
the open Wilson line operator at a fixed time $x_4$ 
carrying momentum $p$ in the commutative direction as
\beqa
&~& \tilde{P}_\mu(x_4 ,p) \label{openWilsonlinesdef2} \\
&\equiv&  \frac{1}{NL} 
\sum_{x_3} e^{- i p x_3 }
\tr \Bigl( V_\mu(x_3,x_4) ^2 \Bigr) \nonumber
\eeqa
for $\mu=1,2$.
We have chosen the power of $V_\mu$ to be the smallest
even number so that the operator couples most effectively
to the Nambu-Goldstone mode associated with the 
spontaneous breakdown of the U(1)$^2$ symmetry.
Then we define the two-point
correlation function of the open Wilson lines
\beq
\tilde{C}_p (\tau) \equiv \frac{1}{2} \sum_{\mu=1}^2 \sum_{x_4}
\Bigl\langle \tilde{P}_\mu(x_4 , p)^{*} \cdot
 \tilde{P}_\mu(x_4 + \tau , p)  \Bigr\rangle
\label{2-pt-cor2}
\eeq
with a separation $\tau$ in the temporal direction.
As in the symmetric phase,
we also consider the case with the 
roles of $x_3$ and $x_4$ exchanged, and take an average over the two
cases to increase the statistics.

We measure the two-point correlation function
and extract the energy at each momentum $p$.
The result is shown in figure \ref{eps:dispersion-broken}.
It is consistent with the expected massless behavior
$E= p$. The discrepancies observed 
at both ends of the spectrum may be interpreted as
finite volume effects and finite lattice-spacing effects, respectively,
and they tend to disappear as $N$ increases.

In the 3d NC $\lambda\phi^4$ theory 
studied in ref.\ \cite{Bietenholz:2004xs},
we only have pseudo Nambu-Goldstone modes,
since the translational symmetry (which is spontaneously broken)
is discretized on the lattice.
The corresponding energy therefore vanishes only in the continuum limit.
In the present case of gauge theory, 
the translational symmetry is enhanced to the continuous U(1)$^2$
symmetry even on the lattice.
Therefore, the energy corresponding to the Nambu-Goldstone mode 
{\em at zero momentum} is exactly zero even
before taking the continuum limit.
This motivated us to introduce a non-zero momentum component $p$
in the commutative direction, which makes the energy non-zero.
(Note also that for $p=0$, we need to subtract the disconnected part, 
which amplifies numerical uncertainties.)
The Nambu-Goldstone mode can still be identified by studying the
dispersion relation.

\section{Summary and discussions}
\label{sec:summary}

In this paper we studied four-dimensional gauge theory
in NC geometry from first principles
based on its lattice formulation.
In particular we clarified the fate of the tachyonic
instability encountered in perturbative calculations.
The lattice formulation is suited for such
a study since the IR singularity responsible for the
instability is regularized in a star-gauge invariant manner,
and we can trace the behavior of the system as the regularization
is removed. This revealed the existence of a first order phase
transition associated with the spontaneous breakdown of 
the U(1)$^2$ symmetry, which includes the translational symmetry
in the NC directions as a discrete subgroup.

In the weak coupling symmetric phase, 
we studied the dispersion relation
and confirmed the presence of an IR singularity.
This IR singularity prevents us from taking the continuum
limit in the symmetric phase. 
While we are able to observe scaling up to some finite $N$,
we cannot increase $N$ further since the energy of the lowest momentum
mode vanishes,
and the vacuum becomes unstable.
In the broken phase, however, we provided evidence
for a sensible continuum limit.
The phase is characterized by the condensation of open Wilson lines,
which represent the ``tachyon'' in the unstable perturbative vacuum.
%
By studying the dispersion relation in the broken phase,
we confirmed the appearance of the Nambu-Goldstone mode
associated with the spontaneous breakdown of the U(1)$^2$ symmetry.

By measuring the eigenvalue distribution of $V_\mu$ in the NC
directions, we find that the dynamical space in the NC directions has
shrunk, but it has a finite {\em physical} extent 
in the continuum limit.
An analogous first order phase transition
is found in gauge theories on fuzzy manifolds \cite{ABNN}, where
the fuzzy manifolds collapse at sufficiently large couplings.
The instability in those cases is due to the uniform condensation 
of a scalar field on the fuzzy manifold\footnote{The 
fuzzy manifolds can be stabilized by adding
a sufficiently large mass to the scalar field \cite{ydri}.
}.
The phenomenon that the space-time itself becomes a dynamical object
is characteristic to gauge theories on NC geometry.
This should be closely related to
the dynamical compactification of extra dimensions\footnote{As a 
closely related line of research,
see refs.\ \cite{CSDR},
which use fuzzy spheres for compactified dimensions.
}
in string theory \cite{SSB,related}
based on the IKKT matrix model \cite{Ishibashi:1996xs}.
Let us recall that the NC geometry appears 
in string theory as 
a result of introducing a background tensor field \cite{String}.
Our conclusion therefore suggests that
the dynamical compactification in string theory
may be associated with the spontaneous generation of
the background tensor field in six dimensions.
This is reminiscent of 
the spontaneous magnetization in
4d non-Abelian gauge theory at finite temperature 
\cite{sp-mag}
or in 3d QED \cite{Hosotani}.


%

On the other hand, if we wish to obtain a 
phenomenologically viable {\em 4d model},
we may stay in the symmetric phase
by keeping the UV cutoff finite and view the NC gauge theory
as an effective theory of a more fundamental theory.
A $\theta$-deformed dispersion relation
for the photon such as the one displayed in figure \ref{eps:dispersion}
should then have implications \cite{Camel}
on observational data from blazars (highly active galactic nuclei)
\cite{Aha}, which are assumed to emit
bursts of photons simultaneously, covering a broad range of energy.
Such experimental efforts will be intensified in the near future.
For instance, the Gamma-ray Large Area Space Telescope
(GLAST) project \cite{GLAST} is scheduled to be launched in
September 2007 and to monitor gamma rays from 20 MeV up to 1 TeV.
In particular, a relative delay of these photons 
depending on the frequency \cite{galax}
could hint at a NC geometry.
%
%


\section*{Acknowledgments} 

We would like to thank
F.~Hofheinz for his contribution to this work in an early
stage, and H. St\"uben for his help concerning parallel computation.
We are also grateful to
T.~Ishikawa for explaining us the details of his work
\cite{ishikawa-okawa} with M.~Okawa,
and to
E.~Lopez for attracting our attention to several relevant references.
We also thank
A.~Bigarini,
H.~Dorn,
N.~Ishibashi,
S.~Iso,
K.~Kanaya,
H.~Kawai,
Y.~Kitazawa,
T.~Konagaya,
Y.~Taniguchi,
A.~Torrielli,
H.~Umetsu
and
K.~Yoshida
for helpful discussions.
The simulations were performed on the IBM p690 clusters
of the ``Norddeutscher
Verbund f\"ur Hoch- und H\"ochstleistungsrechnen'' (HLRN),
and on the PC clusters
at KEK and Humboldt Universit\"{a}t zu Berlin.
%
J.V.\ thanks for the support by
the ``Deutsche Forschungsgemeinschaft'' (DFG).

\appendix

\section{The algorithm for the Monte Carlo simulation}

In this section we describe the algorithm used
to simulate the model (\ref{action}).
The main part of the simulation is performed by the
heat-bath algorithm generalizing ref.\ \cite{Fabricius:1984wp},
where $V_\mu(z)$ is updated by multiplying a SU($N$) matrix.
In order to implement the integration over the U($N$)
--- instead of SU($N$) --- group manifold appropriately,
we also need to include a Metropolis procedure for
updating $V_\mu(z)$ by multiplying a phase factor\footnote{For 
instance, if we used the heat-bath algorithm alone,
$\det V_\mu (z)$ would not change during the simulation.
The Metropolis procedure would be unnecessary if the
model (\ref{action}) were an SU($N$) gauge theory instead of U($N$).
The difference of U($N$) and SU($N$) is irrelevant
in the planar limit, but not necessarily
in the double scaling limit.}.
These procedures, described in the following two subsections,
define ``one sweep'' in our simulation.
Measurements have been performed every 50 sweeps.
The number of configurations used to obtain an ensemble average
is 1980, 1940, 420, 696, 830
for $N=25,35,45,55,65$, respectively, for the sets of parameters
in table \ref{beta-finetuning}.



\subsection{Heat-bath algorithm for multiplying a SU($N$) matrix}

Since two terms, $S_{\rm NC}$ and $S_{\rm mixed}$,
make the action (\ref{action}) non-linear 
with respect to $V_\mu(z)$ for each $\mu$ and $z$,
we cannot apply the heat-bath method \cite{Creutz:1980zw} as it stands.
Note, however, that $S_{\rm NC}$ is nothing but the action 
for the twisted Eguchi-Kawai model for each $z$, which can be
linearized following ref.\ \cite{Fabricius:1984wp}.
Namely we introduce an auxiliary field $Q_{12}(z)$, which is a
general complex $N\times N$ matrix, and consider the action
\beqa
S'_{\rm NC} &=& 
N \beta \sum_{z} \, 
\left[
\tr  \Bigl\{ Q_{12}(z)^\dag Q_{12}(z) \Bigr\} \right. \nonumber \\
&~&  \left.
- 2  \, {\rm Re} \, \tr \left\{ Q_{12}(z)^\dag 
\Bigl( t \, V_1 (z) \, V_2 (z) + 
t^* \, V_2 (z) \, V_1 (z) \Bigr) \right\} \right]  \ ,
\eeqa
where $t$ is a square root
of $ {\cal Z}_{12}$.
Completing the squares
and integrating out the auxiliary field $Q_{12}(z)$,
we retrieve $S_{\rm NC}$.

We use a similar trick to linearize $S_{\rm mixed}$ in (\ref{action}).
Namely we introduce the auxiliary fields 
$Q_{13}(z)$, $Q_{14}(z)$, $Q_{23}(z)$ and $Q_{24}(z)$,
each of which is a general complex $N\times N$ matrix, 
and consider the action
\beqa
S'_{\rm mixed} &=& 
N \beta \sum_z \sum_{\mu=1}^2 \sum_{\nu=3}^4 \, 
\left[ \tr \,  \Bigl(Q_{\mu\nu}(z)^\dagger Q_{\mu\nu}(z)\Bigr)
 \right.  \nonumber \\
&~& \left. - 2 \, {\rm Re} \, 
\tr \, \left\{ Q_{\mu\nu}(z)^\dag
\Bigl( V_\mu(z) V_\nu(z) + V_\nu(z) V_\mu(z+a \hat{\nu})
\Bigr)\right\}  \right]  \ .
\eeqa
Completing the squares
and integrating out the auxiliary fields,
we retrieve $S_{\rm mixed}$.

Since the new action
\beq
S' = S'_{\rm NC} + S_{\rm com} + S'_{\rm mixed}
\label{new-action}
\eeq
is linear with respect to $V_\mu (z)$,
we can use the heat-bath algorithm to update $V_\mu (z)$
by multiplying a matrix in one of the 
$N(N-1)/2$ SU(2) subgroups of the SU($N$) \cite{Cabibbo:1982zn}. 
We repeat this procedure for all the independent SU(2) subgroups.
The auxiliary fields can be easily updated
by generating gaussian variables and shifting them appropriately
depending on the $V_\mu(z)$ field.


\subsection{Metropolis algorithm for multiplying a phase factor}
\label{UNvsSUN}

The updating procedure for rotating the phase of $V_\mu (z)$
is implemented using the Metropolis algorithm.
It turns out that we can increase the acceptance rate by
rotating the phase of the auxiliary field $Q_{\mu\nu}(z)$ 
in a covariant manner as
\beq
\left\{\begin{array}{lcl}
V_\mu(z) & \rightarrow & \ee^{i \alpha_\mu (z)}V_\mu(z) \\
Q_{12}(z) & \rightarrow & \ee^{i \alpha_\mu (z)} Q_{12}(z) \ ,  \\
\end{array}\right.    \quad \quad
\left\{\begin{array}{lcl}
V_\nu(z) & \rightarrow & \ee^{i \alpha_\nu (z)}V_\nu(z) \\
Q_{\mu\nu}(z) & \rightarrow & \ee^{i \alpha_\nu (z)} Q_{\mu\nu}(z) 
\ , \\
\end{array}\right. 
\label{Qrotate}
\eeq
where $\mu=1,2$ and $\nu=3,4$. (No sum is taken in (\ref{Qrotate})
even if an index appears in a term twice.)
At each step, the transformation parameter $\alpha_\mu (z)$
is non-zero for particular $\mu$ and $z$,
and it is given by
a uniform random number within the range
$[0,\frac{2\pi\epsilon}{N}]$, where $\epsilon = 0.1$ was chosen
to keep the acceptance rate reasonably high (e.g., 60\%).
We accept the trial configuration with the probability
$\min (1, \ee^{-\Delta S'})$, $\Delta S' \equiv S'_1
- S' _0$, where $S'_1$ and $S' _0$
are the action evaluated for the trial configuration and the 
present configuration, respectively.
Note that $\Delta S'$ comes solely from
$S'_{\rm mixed}$ ($S'_{\rm com}$) 
when updating $V_{\mu}(z)$ with $\mu=1,2$ ($\mu=3,4$)
thanks to the phase rotation of the auxiliary field (\ref{Qrotate}).
We repeat this procedure for all choices of $\mu$ and $z$.

\end{document}